\documentclass[journal,final]{IEEEtran}
\usepackage{amsfonts}
\usepackage{amssymb}
\usepackage{amsmath}
\usepackage{dsfont}
\usepackage{graphicx,subfigure}
\usepackage{enumitem}
\usepackage{cite}
\usepackage{bigstrut}
\usepackage{float}
\usepackage{balance}
\usepackage{lipsum}
\usepackage{psfrag}
\usepackage{algorithm}
\usepackage[noend]{algpseudocode}
\usepackage[export]{adjustbox}
\usepackage{array}
\newcolumntype{P}[1]{>{\centering\arraybackslash}p{#1}}
\newtheorem{theorem}{Theorem}

\newtheorem{lemma}{Lemma}

\newtheorem{assumption}{Assumption}

\newtheorem{proposition}{Proposition}

\newcommand\norm[1]{\left\lVert#1\right\rVert_2}
\begin{document}

\title{Wireless Quantized Federated Learning: A Joint Computation and Communication Design}
\author{Pavlos S. Bouzinis, \IEEEmembership{	Graduate Student Member,~IEEE}, Panagiotis D. Diamantoulakis,  \IEEEmembership{Senior Member,~IEEE}, and George K. Karagiannidis,  \IEEEmembership{Fellow,~IEEE}
	\thanks{P. S. Bouzinis, P. Diamantoulakis,  and G.K.  Karagiannidis are with Wireless Communication and Information Processing  Group (WCIP), Department of Electrical and Computer Engineering, Aristotle University of Thessaloniki, Greece, E-mails: \{mpouzinis, padiaman, geokarag\}@auth.gr}
}
\maketitle

\begin{abstract}
	Recently, federated learning (FL) has sparked widespread attention as a promising decentralized machine learning approach which provides privacy and low delay. However, communication bottleneck still constitutes an issue, that needs to be resolved for an efficient deployment of FL over wireless networks. In this paper, we aim to minimize the total convergence time of FL, by quantizing the local model parameters prior to uplink transmission. More specifically, the convergence analysis of the FL algorithm with stochastic quantization is firstly presented, which reveals the impact of the quantization error on the convergence rate. Following that, we jointly optimize the computing, communication resources and number of quantization bits, in order to guarantee minimized convergence time across all global rounds, subject to energy and quantization error requirements, which stem from the convergence analysis. The impact of the quantization error on the convergence time is evaluated and the trade-off among model accuracy and timely execution is revealed. Moreover, the proposed method is shown to result in faster convergence in comparison with baseline schemes. Finally, useful insights for the selection of the quantization error tolerance are provided.
\end{abstract}
\section{Introduction} \IEEEPARstart{F}uture wireless networks are envisioned to support ubiquitous artificial intelligent services \cite{letaief}. Conventional machine learning techniques are usually conducted in a centralized manner, where the data are uploaded and processed at a single entity, e.g., a central server \cite{sun2019}. However, the growing computing capabilities of devices have paved the way for realizing distributed learning frameworks. Among the decentralized approaches, federated learning (FL) has shown great promise in preserving data privacy and providing low delay \cite{konevcny_basic}, \cite{McMahan2017}. In FL, users collaboratively build a shared learning model, without exposing their raw data to the server or any other residual participant. The servers' role is to redistribute the global model back to the users, while the whole procedure is repeated until the convergence of the global model. In this manner, FL is inherently privacy-preserving and reduces the communication load. However, the wireless environment imposes some distinctive challenges, owing to the limited wireless resources, unreliable links, etc., which degrade the performance of FL and need to be efficiently addressed \cite{li2020,bouzinis,bouzinis2,lim2020}.
\subsection{Related Works and Motivation}
In the context of wireless networks, several works examined and optimized the performance of FL in various aspects, such as model accuracy, timely execution and energy efficiency. More specifically, in \cite{chen2020}, a joint learning and communication framework was proposed, in order to minimize the training loss 
in the presence of packet errors, while in \cite{yang2020} the objective was to minimize the total energy consumption by optimizing both the computation and communication resources. Moreover in \cite{chen2020convergence}, authors focus on minimizing the convergence time of the FL process by efficiently scheduling the participating devices, while in \cite{Wan2021}, the convergence time was minimized by considering both the impact of training and communication. In \cite{shi2020}, the FL global loss function was minimized under total convergence time constraints, by jointly allocating the bandwidth and scheduling the users, while in \cite{yang2019}, the impact of various scheduling policies on the FL convergence rate was examined.
\par
Due to communication bottleneck incurred by the limited bandwidth and the large size of the local training parameters, gradient compression techniques, such as quantization, have been proposed to further improve the communication efficiency in FL \cite{konevcny1,fedpaq,caldas2018}. Through this technique, a quantized version of the local model is being transmitted to the server, aiming to achieve faster communication without deteriorating the FL model accuracy. In this direction, authors in \cite{zheng2020}, examined quantization schemes for the uplink and downlink communication in FL and rigorously proved the respective convergence rate upper bounds. In \cite{amiri2020q}, a lossy FL scheme is introduced where both global and local updates are quantized before being transmitted, while the convergence behavior is analyzed. It is concluded that the performance loss compared to an ideal model, is negligible. Moreover, in \cite{chang2020}, the minimization of the convergence bound was performed, in a multiple access channel scenario of FL, while an efficient utilization of the quantization levels was proposed. In \cite{wang2021}, the quantization error was minimized subject to uplink transmission delay and outage constraints per round, in a wireless environment. In \cite{zhu2020}, authors proposed a one-bit quantization scheme for over-the-air FL, while they also examined the convergence rate under fading channels.
\par
At this point, we clarify that none of the aforementioned works \cite{zheng2020,amiri2020q,chang2020,wang2021,zhu2020}, who applied quantization methods, focused on minimizing the total convergence time of the FL, which is a metric of paramount importance for meeting the low latency requirements of the next generation of wireless networks. Regarding the works who performed optimization methods, \cite{chang2020} and \cite{wang2021} focused on minimizing the quantization error without jointly considering the computation, the communication resources, as well as the quantization policy. Specifically, \cite{chang2020} did not consider the wireless factors, while in \cite{wang2021}, the local computation time and subsequently the adjustment of users' local CPU frequency were neglected. Therefore, guarantying fast convergence and high model accuracy of FL, when applying quantization techniques, by efficiently utilizing the available computation, radio resources and quantization strategy, is an issue that has not been yet resolved. Furthermore, the trade-off among model accuracy and fast convergence, owing to the local updates' quantization, is not well-examined in the previous works, where the convergence rate is mainly investigated with regards to the global communication rounds and not the total evolution of time. Specifically, a large number of quantization bits may increase precision, in the expense of slow communication with the server per FL round, since the training parameters' size also increases and leads to higher transmission delay. On the contrary, smaller quantization level may lead to slightly decreased model performance, though, with lower delay per round and potentially faster convergence. However, low precision updates may be communication efficient in terms of latency per round, but require increased number of communication rounds until convergence. Therefore, it is not evident that the utilization of few quantization bits can always lead to faster convergence, with respect to the evolution of time. The aforementioned fact is not clearly presented in the literature and needs further investigation.
\subsection{Contributions}
Driven by the aforementioned considerations, we study the quantization of the local model parameters of each user and aim to minimize the total convergence time of FL, under energy consumption and quantization error constraints. The latter aims to retain the model accuracy in desirable levels. Moreover, the considered minimization is conducted with respect to the unit of time, and not purely the evolution of global FL rounds, since the wireless factors, the available resources and the quantization policy affect the duration of each round, which is critical for the evaluation of the total convergence time.
\par
The main contributions of this paper are summarized as:
\begin{itemize}
	\item We present a rigorous convergence analysis of the FL process by considering stochastic quantization of the local parameters. The impact of the quantization error on the convergence bound is revealed, while useful insights for the optimality gap are provided. Moreover, the derivation of the resulted bound does not enforce the quantization bits to take specific values, which is requisite for performing optimization methods subject to various constraints, and thus, allowing the adjustment of the quantization bits accordingly.
	\item We minimize the total convergence time of the FL across all global rounds, subject to energy consumption and quantization error constraints. The latter have resulted from the convergence analysis and aim to guarantee sufficiently high precision and subsequently high model accuracy. To this end, we jointly optimize the computation and communication resources, as well as the number of quantization bits of each user. After some mathematical manipulations, the resulted convex problem is solved with the \textit{Lagrange dual decomposition} and closed-form solutions are derived, in terms of the Lagrange multipliers (LMs).
	\item Through simulations, the performance of local parameters' quantization is evaluated. Specifically, we investigate the impact of the quantization error tolerance on the convergence time and model accuracy, while the trade-off among model accuracy and fast convergence is exhibited. In addition to this, it is shown that low precision quantization cannot always achieve fast convergence. Moreover, numerical results validate the effectiveness of the proposed optimization towards minimizing the convergence time, in comparison with baseline schemes. Finally, driven by the theoretical convergence analysis, we study the effects of decaying the quantization error tolerance along with the evolution of the training, instead of keeping it constant. The simulation results corroborate the effectiveness of this approach, which demonstrates increased convergence rate without model accuracy degradation. In essence, this observation coincides with the concept of ``later-is-better" \cite{shen2021}, which implies that reserving FL-related resources in the early stages of the training process and spending them in the final stages, may be beneficial for the performance. 
\end{itemize}
\section{System model}
\subsection{FL model}
We consider a wireless FL system, consisting of $N$ users, indexed as $n\in\mathcal{N}=\{1,2,...,N\}$ and a base station (BS) co-located with a server, while hereinafter we use the terms BS and server interchangeably. Each user $n$, poses a local dataset $\mathcal{D}^{\mathrm{L}}_n=\{\boldsymbol{x}_{n,k},y_{n,i}\}^{D^{\mathrm{L}}_n}_{k=1}$, where $D^{\mathrm{L}}_n=\vert\mathcal{D}^{\mathrm{L}}_n\vert$, $\boldsymbol{x}_{n,k}$ is the $k$-th input data vector of user $n$, while $y_{n,k}$ is the corresponding output. The whole dataset is denoted as $\mathcal{D}=\underset{n \in \mathcal{N}}{\cup }\mathcal{D}^{\mathrm{L}}_n$, while the size of all training data is  $D=\sum_{n=1}^{N}D^{\mathrm{L}}_n$. The loss function of user $n$, is defined as \cite{lim2020}
\begin{equation}
F_n(\boldsymbol{w}) \triangleq \frac{1}{D^{\mathrm{L}}_n}\sum_{k \in \mathcal{D}^{\mathrm{L}}_n}\phi(\boldsymbol{w},\boldsymbol{x}_{n,k},y_{n,k}), \quad \forall n \in \mathcal{N},
\end{equation}
where $\phi(\boldsymbol{w},\boldsymbol{x}_{n,k},y_{n,k})$ captures the error of the $d$-dimensional model parameter
$\boldsymbol{w}\in\mathbb{R}^d$ for the input-output pair $\{\boldsymbol{x}_{n,k},y_{n,k}\}$. The goal of the training process is to find the global parameter $\boldsymbol{w}$, which minimizes the loss function on the whole dataset
\begin{equation}
	F(\boldsymbol{w})=\sum_{n=1}^{N}p_nF_n(\boldsymbol{w}),
\end{equation}
where $p_n=\frac{D^{\mathrm{L}}_n}{D}$, i.e., to find $\boldsymbol{w}^*= \underset{\boldsymbol{w}}{\mathrm{argmin}}\,F(\boldsymbol{w}).$
\par We assume that the whole FL process consists of $T$ global rounds, denoted as $t\in\mathcal{T}=\{0,...,T-1\}$. During the $t$
-th global round, each user receives the global parameter $\boldsymbol{w}(t)$ from the server, and performs $\tau$ steps of the stochastic gradient descent (SGD) method. The $i$-th step of SGD, $\forall n\in \mathcal{N}$, is given as
\begin{equation}
\boldsymbol{w}^i_n(t)=\boldsymbol{w}^{i-1}_n(t)-\eta(t)\nabla F_n(\boldsymbol{w}^{i-1}_n(t),\boldsymbol{\xi}^{i-1}_n(t)), \, \, i=1,...,\tau \, ,
\end{equation}
where $\boldsymbol{w}^0_n(t)\triangleq \boldsymbol{w}(t)$. Moreover, $\eta(t)$ represents the learning rate, while $\boldsymbol{\xi}^{i-1}_n(t)\subseteq \mathcal{D}^{\mathrm{L}}_n$ is a mini-batch, which is sampled uniformly at random from the local dataset $\mathcal{D}^{\mathrm{L}}_n$ of user $n$. Therefore, it holds  $\mathbb{E}[\nabla F_n(\boldsymbol{w}^{i-1}_n(t),\boldsymbol{\xi}^{i-1}_n(t))]=\nabla F_n(\boldsymbol{w}^{i-1}_n(t))$, where the expectation is taken with respect to the randomness of the stochastic gradient function. Furthermore, we assume that at the first global round, the server initializes $\boldsymbol{w}(0)$. After terminating the local training, user $n$ transmits the weight differential to the server, given as
\begin{equation}
\begin{split}
	\Delta \boldsymbol{w}_n(t) &=\boldsymbol{w}^{\tau}_n(t)-\boldsymbol{w}^0_n(t)=\boldsymbol{w}^{\tau}_n(t)-\boldsymbol{w}(t)\\
	&=-\eta(t)\sum_{i=1}^{\tau}\nabla F_n(\boldsymbol{w}^{i-1}_n(t),\boldsymbol{\xi}^{i-1}_n(t)).
	\label{DW_n}
\end{split}
\end{equation}
The selection of transmitting the weight differential instead of simply transmitting the latest local weight $\boldsymbol{w}^{\tau}_n(t)$, is related with the quantization scheme that will be used and discussed later on this work.
 Following that, the global model at the server's side, in round $t$, is updated as follows
\begin{equation}
	\boldsymbol{w}(t+1)=\boldsymbol{w}(t)+\sum_{n=1}^{N}p_n\Delta \boldsymbol{w}_n(t).
\end{equation}
At last, the global model is broadcast to the devices, while the whole process is repeated for $T$ rounds, until the convergence of the global model.
\subsection{Quantization model}
As mentioned previously, at time step $t$, each user $n \in \mathcal{N}$ calculates its local model $\Delta \boldsymbol{w}_n(t)=(\Delta w_{n,1}(t),...,\Delta w_{n,d}(t))^{\top}$. In order to prevent a wasteful overuse of resources, we assume that users send to the server a quantized version of $\Delta \boldsymbol{w}_n(t)$, which is denoted as $Q(\Delta \boldsymbol{w}_n(t))$, where $Q(\cdot)$ denotes the quantization function. Therefore, the global model update at the server, is actually given as
\begin{equation}
\boldsymbol{w}(t+1)=\boldsymbol{w}(t)+\sum_{n=1}^{N}p_nQ(\Delta \boldsymbol{w}_n(t)).
\vspace{0.2cm}
\end{equation}
We also highlight that in \cite{zheng2020} it was shown that by transmitting the weight differential $\Delta \boldsymbol{w}_n(t)$, a faster convergence is achieved, while it is also adopted in our work.
Following that, for each element $j\in\{1,2,...,d\}$ of $\Delta \boldsymbol{w}_n(t)$, it holds $\vert \Delta w_{n,j}(t)\vert \in [\Delta w^{\mathrm{min}}_n(t),\Delta w^{\mathrm{max}}_n(t)]$, where $\Delta w^{\mathrm{min}}_n(t)\triangleq\mathrm{min}\{\vert\Delta \boldsymbol{w}_n(t)\vert\}$ and $\Delta w^{\mathrm{max}}_n(t)\triangleq\mathrm{max}\{\vert\Delta \boldsymbol{w}_n(t)\vert\}$.  Moreover, we assume that $\Delta w_{n,j}(t)$ is quantized according to the stochastic quantization method \cite{zheng2020}. That is, with $B_n(t)$ quantization bits, user $n$ can divide the interval $[\Delta w^{\mathrm{min}}_{n}(t),\Delta w^{\mathrm{max}}_{n}(t)]$ into the following $\varsigma$ intervals:  $I_1=[s_0,s_1], \, I_2=[s_1,s_2],..., I_{\varsigma}=[s_{\varsigma-1},s_{\varsigma}]$, where $\varsigma = 2^{B_n(t)}-1$ and
\begin{equation}
s_k=\Delta w^{\mathrm{min}}_{n}(t)+k\frac{\Delta w^{\mathrm{max}}_{n}(t)-\Delta w^{\mathrm{min}}_{n}(t)}{2^{B_n(t)}-1}, 
\end{equation}
where $k=0,1,...,2^{B_n(t)}-1.$
Therefore, if the parameter $\Delta w_{n,j}(t)$ falls into the interval $I_k$, it will be quantized as
\begin{equation}
Q(\Delta w_{n,j}(t))= \begin{cases} s_{k-1}\cdot\mathrm{sign}(\Delta w_{n,j}(t)), \quad \mathrm{w.p.} \, \tfrac{s_k-\vert \Delta w_{n,j}(t)\vert}{s_k-s_{k-1}}\\
s_k\cdot\mathrm{sign}(\Delta w_{n,j}(t)), \quad \mathrm{w.p.} \, \tfrac{\vert \Delta w_{n,j}(t)\vert-s_{k-1}}{s_k-s_{k-1}}
\end{cases}
\end{equation}
where w.p. stands for ``with probability". Furthermore,
the overall $d$-dimensional quantized local model is denoted as 
\begin{equation}
Q(\Delta \boldsymbol{w}_n(t))=(Q(\Delta w_{n,1}(t)),...,Q(\Delta w_{n,d}(t)))^\top, \quad \forall n,t.
\end{equation}
while its size is given by
\begin{equation}
S_n(t)=d(B_n(t)+1)+m \quad \mathrm{(bits)},
\end{equation}
since each element of the quantized model vector is represented by $B_n(t)$ bits plus one bit for the sign specification. Furthermore, $m$ bits are needed to specify the values of $\Delta w^{\mathrm{max}}_{n}(t)$ and $\Delta w^{\mathrm{min}}_{n}(t)$.
\begin{figure}[t!]
	\centering
	\includegraphics[width=0.85\linewidth]{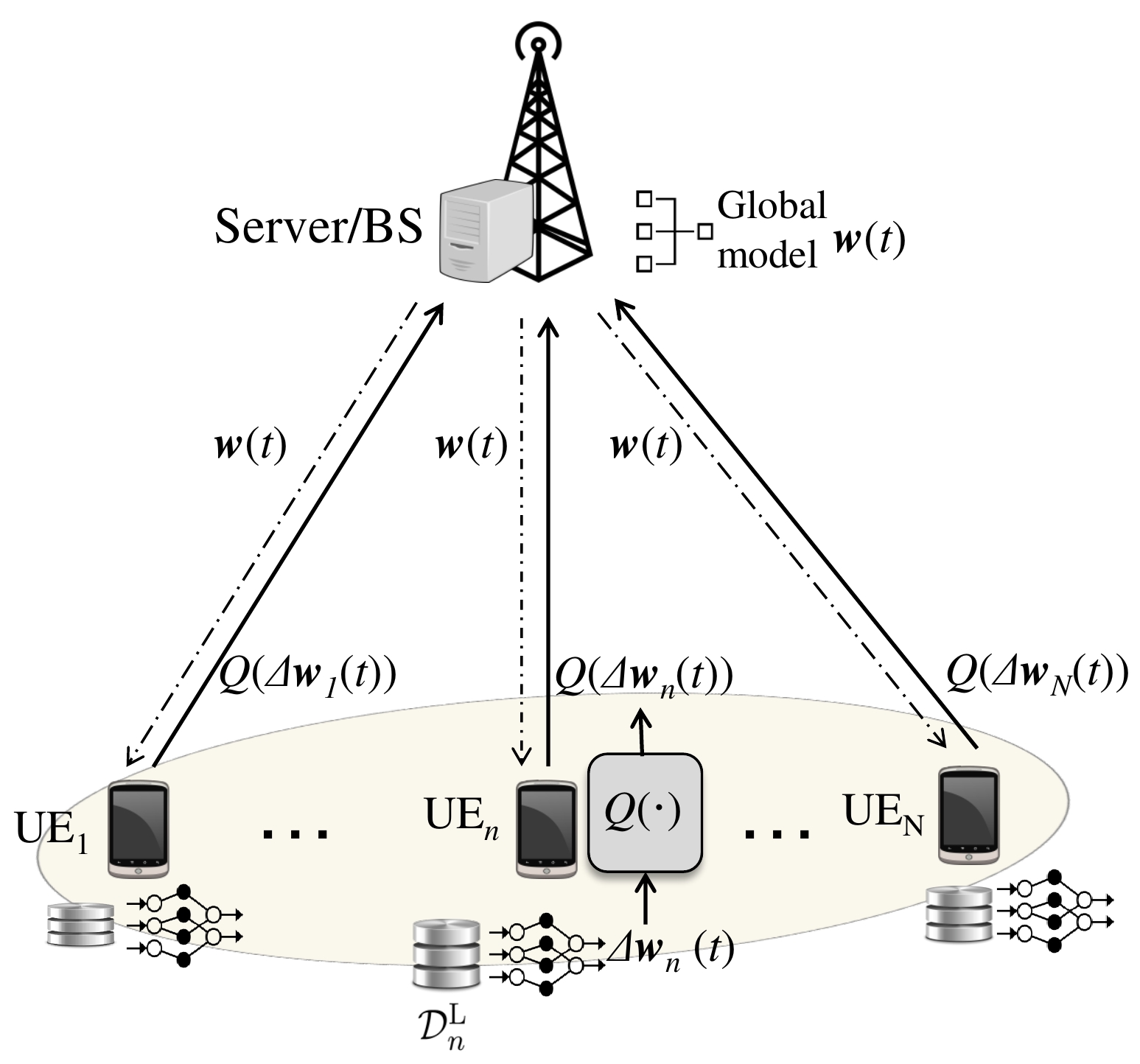}
	\caption{Federated learning with local model quantization.}
	\label{sm}
\end{figure}
\subsection{Computation and Communication model}
During the $t$-th global round, the server is broadcasting the global model $\boldsymbol{w}(t)$ to all users. We consider that the downlink transmission latency is negligible, since the transmit power of the BS is much larger than that of the devices, while also the same message is broadcast to all users. In addition to this, we assume that the downlink is error-free, i.e., all users correctly receive the global model updates. In the continue, for ease of presentation, we slightly abuse the notation by dropping  $t$, while the following expressions could refer to any arbitrary round.
The dedicated time for local computations by each device $n$, for $\tau$ SGD steps, in order to generate the local model, is given as
\begin{equation}
	l^\mathrm{c}_n=\tau\frac{c_nD_n}{f_n}, \quad \forall n \in \mathcal{N},
\end{equation}
where $f_n$ is the CPU cycle frequency of user $n$, $D_n$ is the size of the mini-batch (in bits), while $c_n$ denotes the number of CPU cycles for user $n$ to perform one sample
of data during the local model training. The corresponding energy consumption for the local computations, is given as
\vspace{0.1cm}
\begin{equation}
	E^\mathrm{c}_n=\tau\zeta c_nD_nf^2_n, \quad \forall n \in \mathcal{N},\vspace{0.1cm}
\end{equation}
where $\zeta$ is a constant parameter related with the hardware architecture of device $n$.
\par
\begin{figure}[t!]
	\centering
	\includegraphics[width=0.7\linewidth]{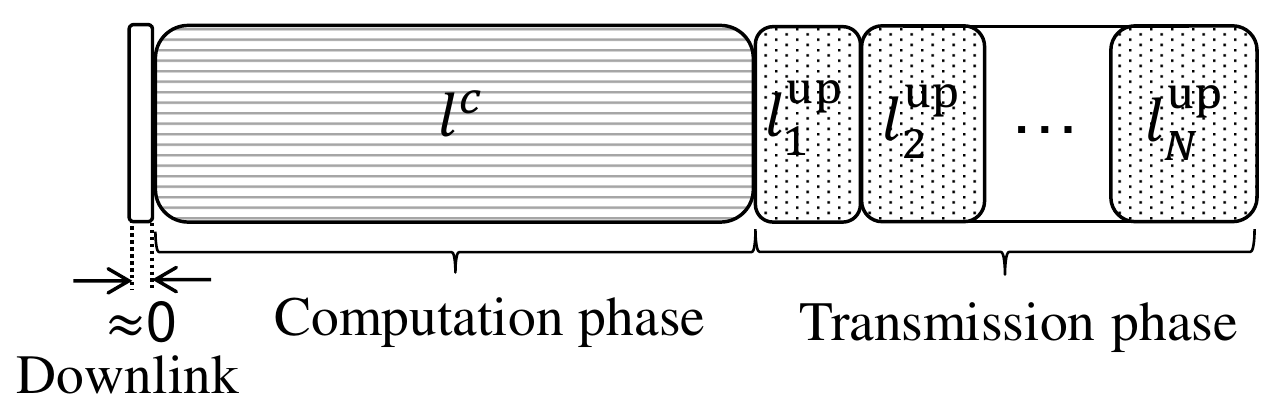}
	\caption{Duration of a global FL round.}
	\label{delayFig}
\end{figure}

 Following the local training, each device uploads the quantized training parameters, $Q(\Delta \boldsymbol{w}_n(t))$, to the BS. Similarly to \cite{tran2019}, we assume that the considered transmission is carried out via time-division multiple access (TDMA), while this choice is not restrictive, since other schemes such as frequency-division multiple access (FDMA) can also be applied. To successfully upload the training parameters within $l^{\mathrm{up}}_n$ uplink time duration, the $n$-th user should satisfy the condition
\vspace{0.1cm}
\begin{equation}
	l^{\mathrm{up}}_nW\log_2\left(1+\frac{g_nE_n}{l^{\mathrm{up}}_nWN_0}\right)\geq S_n, \quad \forall n \in \mathcal{N},\vspace{0.1cm}
\end{equation}
where $W$ is the available bandwidth, $E_n$ is the transmit energy, $S_n$ is the size of the quantized training parameters and $N_0$ is the power of the additive white Gaussian noise (AWGN). Moreover, $g_n=\vert h_n \vert^2d_n^{-\beta}$ denotes the channel gain, where the complex random variable $h_n \sim \mathcal{CN}(0,1)$
is the small scale fading, $d_n$ is the distance between user
$n$ and the BS and $\beta$ is the path loss exponent. Moreover, we assume that the channel gain is quasi-static and stays unchanged during a global round, while we also consider perfect CSI both in BS's and users' side.
Following that, we assume that the parameter transmission phase begins after the termination of the local computations phase by each user. Therefore, for the local computation duration $l^{\mathrm{c}}$, it holds that
\begin{equation}\label{lc}
	l^{\mathrm{c}}\geq l^\mathrm{c}_n=\tau\frac{c_nD_n}{f_n}, \quad \forall n \in \mathcal{N}.
\end{equation}
Since all users should terminate the computation and uplink transmission phases, so as the server can receive each local model and subsequently update the global model, the total duration of a FL global round is given as
\begin{equation}
	l^{\mathrm{r}}=l^{\mathrm{c}}+\sum_{n=1}^{N}l^{\mathrm{up}}_n,
\end{equation}
which is the sum of the computation latency and the transmission latency among all users and is depicted in Fig. \ref{delayFig}.

\section{Convergence Analysis}
In this section we investigate the convergence behavior of FL with stochastic quantization ans stochastic gradient function. Firstly, with regards to the stochastic quantization scheme, we proceed to the formulation of the following lemma:
\begin{lemma}
	$Q(\Delta \boldsymbol{w}_n(t))$ is an unbiased estimator of $\Delta \boldsymbol{w}_n(t)$, i.e.,
	\begin{equation}
	\mathbb{E}\left[Q(\Delta \boldsymbol{w}_n(t))\right]=\Delta \boldsymbol{w}_n(t),
	\end{equation}
	while it also holds that
	\begin{equation}
	\mathbb{E}\left[\norm{Q(\Delta \boldsymbol{w}_n(t))-\Delta \boldsymbol{w}_n(t)}^2\right]\leq\frac{\delta^2_n(t)}{\left(2^{B_n(t)}-1\right)^2}\triangleq J^2_n(t),
	\end{equation}
	where $\delta_n(t)\triangleq \sqrt{\frac{d}{4}\left(\Delta w^{\mathrm{max}}_{n}(t)-\Delta w^{\mathrm{min}}_{n}(t)\right)^2}$.
\end{lemma}
\vspace{0.2cm}
\begin{IEEEproof}
	The proof can be found in \cite{zheng2020},\cite{wang2021}.
	\vspace{0.2cm}
\end{IEEEproof}
\par Next, we make the following common assumptions for the functions $F_1,F_2,...,F_N$, in order to facilitate the convergence analysis \cite{li2019}:
\vspace{0.1cm}
\begin{assumption}
	$F_n, \, \forall n \in \mathcal{N},$ are all L-smooth, i.e.,  $\forall\boldsymbol{w}',\boldsymbol{w}\in\mathbb{R}^d$: $F_n(\boldsymbol{w}') \leq F_n(\boldsymbol{w})+\langle\boldsymbol{w}'-\boldsymbol{w},\nabla F_n(\boldsymbol{w})\rangle+\frac{L}{2}\norm{\boldsymbol{w}'-\boldsymbol{w}}^2$. 
\end{assumption}
\vspace{0.2cm}
\begin{assumption}
	$F_n, \, \forall n \in \mathcal{N},$ are all $\mu$-strongly convex, i.e., $\forall\boldsymbol{w}',\boldsymbol{w}\in\mathbb{R}^d$: $F_n(\boldsymbol{w}') \geq F_n(\boldsymbol{w})+\langle\boldsymbol{w}'-\boldsymbol{w},\nabla F_n(\boldsymbol{w})\rangle+\frac{\mu}{2}\norm{\boldsymbol{w}'-\boldsymbol{w}}^2$. 
\end{assumption}
\vspace{0.2cm}
\begin{assumption}
 The expected squared norm of stochastic gradients is uniformly bounded $\forall n,t,i$, i.e., $\mathbb{E}\left[\norm{\nabla F_n(\boldsymbol{w}^i_n(t),\boldsymbol{\xi}^i_n(t))}^2\right]\leq G^2$.
\end{assumption}
\vspace{0.2cm}
\begin{assumption}
	The variance of stochastic gradients in each user is bounded $\forall n,t,i$, i.e., $\mathbb{E}\left[\norm{\nabla F_n(\boldsymbol{w}^i_n(t),\boldsymbol{\xi}^i_n(t)-\nabla F_n(\boldsymbol{w}^i_n(t))}^2\right]\leq \sigma^2_n$.
\end{assumption}
\vspace{0.1cm}
Moreover, we define $\Gamma \triangleq F(\boldsymbol{w}^*)-\sum_{n=1}^{N}p_nF^*_n$, where $F^*_n$ denotes the minimum value of $F_n(\cdot)$, while $\Gamma$ quantifies the degree of NON-IID, among users' datasets. Taking these into account, we introduce the following theorem:
\vspace{0.1cm}
\begin{theorem} Let Assumptions 1 to 4 hold. By selecting a diminishing learning rate $\eta(t)=\frac{2}{\mu(\gamma+t)}$ and $\gamma>\mathrm{max}\big\{2,\frac{2}{\mu},\frac{L}{\mu}\big\}$, the upper bound of $\mathbb{E}\left[F(\boldsymbol{w}(T))-F(\boldsymbol{w}^*)\right]$ is given by
	\vspace{0.1cm}
	\begin{equation}\label{TH1}
	\begin{split}
	&\mathbb{E}\left[F(\boldsymbol{w}(T))\right]-F(\boldsymbol{w}^*)\leq\\[2pt]
	& \qquad \quad \frac{L}{2}\frac{1}{\gamma + T}\left(\frac{4U}{\mu^2}+\gamma\mathbb{E}\left[\norm{\boldsymbol{w}(0)-\boldsymbol{w}^*}^2\right]\right) \\[5pt]
	&\qquad + \underbrace{\frac{L}{2}\sum_{j=0}^{T-1}\Bigg[\sum_{n=1}^{N}p_n\frac{\delta^2_n(j)}{\left(2^{B_n(j)}-1\right)^2}\prod_{i=j+1}^{T-1}\left(1-\frac{2}{\gamma+i}\right)\Bigg]}_{\text{Impact of the quantization error on the convergence}},
	\end{split}
		\end{equation}
where
\begin{equation}
U=\tau^2\sum_{n=1}^{N}\sigma^2_n+\tau G^2+ 2L\tau^2\Gamma
+ (\mu+2)\frac{\tau(\tau-1)(2\tau-1)}{6}G^2,
\end{equation}
while the expectation is taken with respect to the stochastic gradient function and the stochastic quantization scheme.
\end{theorem}
\begin{IEEEproof}
	See Appendix A.
\end{IEEEproof}

As one can observe, for large $T$, the first term of the upper bound tends to zero. However, the second term which is related with the quantization error, creates a gap between $\mathbb{E}\left[F(\boldsymbol{w}(T))\right]$ and $F(\boldsymbol{w}^*)$. Inspired by \cite{amiri2020q}, we present the following interesting comment. For small $j$, the term $\prod_{i=j+1}^{T-1}\left(1-\frac{2}{\gamma+i}\right)$ tends to zero, since $1-\frac{2}{\gamma+i}<1$, $\forall i$. Therefore, the effect of the quantization error in the early stages of the training process vanishes over time. Hence, it is discernible that during the early training rounds, the quantization error would not contribute in increasing the optimality gap.
 Nevertheless, in order to further mitigate the impact of the quantization error, an increased number of quantization bits $B_n$ may be selected. However, such choice may result to increased latency during the local parameter transmission phase, while the considered trade-off is studied later. Finally, note that when $B_n(j)\rightarrow \infty, \, \forall n,j$, the optimality gap is zero and the convergence bound of Theorem 1 coincides with that of a lossless FL model.
\par
 At this point, it should be highlighted that in the convergence analysis of \cite{zheng2020}, where a stochastic quantization scheme was also considered, authors concluded that the quantization error does not create an optimality gap, i.e., $\mathbb{E}\left[F(\boldsymbol{w}(T))-F(\boldsymbol{w}^*)\right]$ tends asymptotically to zero. However, in their analysis, they selected specific values for the quantization bits $B_n(t)$, given as a function of the learning rate $\eta(t)$. In opposition to this, in our analysis we do not restrict $B_n(t)$ to take certain values. In this manner, the values of $B_n(t)$ are not being enforced by $\eta(t)$. This fact is of significant importance, since the constraints imposed by the wireless environment, could affect the selection of $B_n(t)$, i.e., it cannot be always feasible or communication efficient to pre-assign specific values to $B_n(t)$ based on the learning rate. The constraints on $B_n(t)$, would be obvious through the optimization problem that is formulated in the subsequent section.
\section{Convergence Time Minimization}
\subsection{Problem Formulation}
Our objective goal is to minimize the total convergence time of the FL task, i.e., the overall latency across all FL rounds, under energy and quantization error constraints, with the latter aiming to retain the optimality gap of the upper bound of $\mathbb{E}\left[F(\boldsymbol{w}(T))\right]-F(\boldsymbol{w}^*)$, at small levels. Note that the upper bound is affected by the quantization error through the term $\sum_{n=1}^{N}p_n\frac{\delta^2_n(t)}{\left(2^{B_n(t)}-1\right)^2}$, \vspace{0.03cm} as concluded in Theorem 1. Therefore, it is obvious that by increasing the number of quantization bits $B_n(t)$, the quantization error decreases. However, this strategy increases the size $S_n(t)$ of the local model parameters and thus, may result in increased transmission latency. Taking this into account, it is important to balance the considered trade-off, among model accuracy and fast convergence. Hence, we formulate the following optimization problem
\vspace{0.3cm}
\begin{equation}
\begin{aligned}
& &  &\underset{l^{\mathrm{c}},\boldsymbol{E},\boldsymbol{B},\boldsymbol{f},\boldsymbol{l}^\mathrm{up}}{\text{min}}  \quad
\text{\emph{{$\sum_{t=0}^{T-1}l^{\mathrm{r}}(t)$}}} \\ 
& & &\textbf{\emph{s.t.}}
\quad \mathrm{C}_{1}: l^{\mathrm{up}}_n(t)W\log_2\left(1+\frac{g_n(t)E_n(t)}{l^{\mathrm{up}}_n(t)WN_0}\right)\\
& & &  \qquad \qquad \geq d(B_n(t)+1)+m, \quad \forall n \in \mathcal{N}, \forall t,\\
& & &\qquad \mathrm{C}_{2}: 	\tau \zeta c_nD_nf^2_n(t)+E_n(t) \leq E^{\mathrm{max}}_n(t), \, \,\forall n \in \mathcal{N}, \forall t,\\
& & & \qquad   \mathrm{C}_3: \sum_{n=1}^{N}p_n\frac{\delta^2_n(t)}{\left(2^{B_n(t)}-1\right)^2} \leq \epsilon(t), \quad \forall t,\\
& & & \qquad   \mathrm{C}_4:  l^{\mathrm{c}}(t) \geq \tau\frac{c_nD_n}{f_n(t)}, \quad \forall n \in \mathcal{N}, \quad \forall t,\\
& & & \qquad   \mathrm{C}_5:0 \leq f_n(t) \leq f^{\mathrm{max}}_n(t), \quad B_n(t) \in \mathbb{Z}_{+}, \quad \forall t,
\end{aligned}\label{primary}
\end{equation}
where $\mathrm{C}_1$ is related with the successful transmission of the local training parameters, $\mathrm{C}_2$ indicates that the dedicated energy both for computation and transmission purposes, cannot exceed the maximum available energy of the $n$-th user at the $t$-th round, i.e., $E^{\mathrm{max}}_n(t)$. Moreover, $\mathrm{C}_3$ implies that the quantization error should not exceed a required tolerance, $\epsilon(t)$, at the respective round. We also clarify that it is reasonable to constrain the quantization error per global FL round $t$, since there is no coupling of the error among different global rounds, as observed in \eqref{TH1}. Finally, $\mathrm{C}_4$ stems from \eqref{lc}, while $f^{\mathrm{max}}_n$ denotes the maximum CPU clock speed of user $n$. Also, note that the quantization bits $B_n(t)$  are positive integers. Moreover, we highlight that the selection of increased number of bits $B_n(t)$, leads to better model precision, which is reflected in $\mathrm{C}_3$ through the selection of the error tolerance $\epsilon(t)$. However, it is observed from $\mathrm{C}_1$ that such policy also increases the transmission delay and subsequently the total convergence time.
\subsection{Proposed Solution}
It should be highlighted that problem  \eqref{primary} is intractable in the current formulation, since at the $t$-th round the channel gains $g(t'),\,\forall t'>t$ are unknown. However, this is not restricting, since we can address this issue by solving the problem round-by-round, in an online fashion.  In addition to this, we relax $B_n \in \mathbb{Z}_{+}$ to $B_n \geq 1, \,\forall n \in \mathcal{N}$. Thus, the  problem in  \eqref{primary} should be solved in each global round, while hereinafter the $t$ notation is dropped for simplicity.

Next, by observing that $\mathrm{C}_4$ is equivalent to: $ f_n \geq \tau\frac{c_nD_n}{l^{\mathrm{c}}}, \, \forall n \in \mathcal{N}$, we introduce the following proposition:
\begin{proposition}
	The optimal $f_n, \forall n \in \mathcal{N}$, satisfy
	\begin{equation}
		f^*_n=\tau\frac{c_nD_n}{l^{\mathrm{c}*}}, \quad \forall n 
	\in \mathcal{N},
	\label{fn_optimal}
	\end{equation}
	with $l^{\mathrm{c*}}\geq a_1 \triangleq \underset{n \in \mathcal{N}}{\mathrm{max}}\left\{\frac{\tau c_nD_n}{f^{\mathrm{max}}_n}\right\}$.
\end{proposition}
\begin{IEEEproof}
	Firstly, by manipulating $\mathrm{C}_4$, it is straightforward to show that $l^\mathrm{c}\geq a_1$. Following that, let consider a known $\bar{l^\mathrm{c}}$. By observing $\mathrm{C}_2$, it is obvious that the selection of larger $f_n$ decreases the value of $E_n$. Moreover, it easy to verify that $l^{\mathrm{up}}_nW\log_2\left(1+\frac{g_nE_n}{l^{\mathrm{up}}_nWN_0}\right)$ in $\mathrm{C}_1$, is an increasing function w.r.t. both $E_n$ and $l^{\mathrm{up}}_n$. Therefore, the selection of smaller $E_n$ will lead to increased $l^{\mathrm{up}}_n$, while the objective is to minimize $l^{\mathrm{up}}_n$. From the aforementioned, $f_n$ should be selected as small as possible, given the local computation duration, concluding to \eqref{fn_optimal}.
\end{IEEEproof}
Proposition 1 implies that the CPU clock speed $f_n$ should be selected in such way that all users terminate the computation phase concurrently. 
By exploiting Proposition 1, the problem can be re-written as
\begin{equation}
\begin{aligned}
& &  &\underset{l^{\mathrm{c}},\boldsymbol{E},\boldsymbol{B},\boldsymbol{l}^\mathrm{up}}{\text{min}} \quad
\text{\emph{{$l^{\mathrm{c}}+\sum_{n=1}^{N}l^{\mathrm{up}}_n$}}} \\ 
& & &\textbf{\emph{s.t.}}
\quad \mathrm{C}_{1}: l^{\mathrm{up}}_nW\log_2\left(1+\frac{g_nE_n}{l^{\mathrm{up}}_nWN_0}\right)\\
& & & \qquad \qquad \geq d(B_n+1)+m, \quad \forall n \in \mathcal{N},\\
& & &\qquad \mathrm{C}_{2}: \frac{\zeta \tau^3c^3_nD^3_n}{{l^{\mathrm{c}}}^2}+E_n \leq E^{\mathrm{max}}_n, \quad \forall n \in \mathcal{N}, \\
& & & \qquad   \mathrm{C}_3: \sum_{n=1}^{N}p_n\frac{\delta^2_n}{\left(2^{B_n}-1\right)^2} \leq \epsilon, \\
& & & \qquad   \mathrm{C}_4:  l^{\mathrm{c}} \geq a_1, \quad E_n,l^{\mathrm{up}}_n \geq 0, \,\forall n \in \mathcal{N},\\
& & & \qquad   \mathrm{C}_5: B_n \geq 1, \,\forall n \in \mathcal{N}.
\end{aligned}\label{primary2}
\end{equation}
It can be easily shown that the problem in \eqref{primary2} is jointly convex w.r.t. all the considered variables, while the proof is omitted due to space limitations.
\par
The problem in \eqref{primary2} will be solved via the \textit{Lagrange dual decomposition}. Firstly, the Lagrangian function can be written as
\begin{equation}
	\begin{split}
	&\mathcal{L}(l^{\mathrm{c}},\boldsymbol{l}^\mathrm{up},\boldsymbol{E},\boldsymbol{B},\boldsymbol{\lambda})=l^{\mathrm{c}}+\sum_{n=1}^{N}l^{\mathrm{up}}_n\\
	& \quad +\sum_{n=1}^{N}\lambda_{1,n}\left(d(B_n+1)+m-l^{\mathrm{up}}_nW\log_2\left(1+\tfrac{g_nE_n}{l^{\mathrm{up}}_nWN_0}\right)\right)\\
	& \quad +\sum_{n=1}^{N}\lambda_{2,n}\left(\frac{\zeta \tau^3c^3_nD^3_n}{{l^{\mathrm{c}}}^2}+E_n - E^{\mathrm{max}}_n\right)\\
	& \quad +\lambda_3\left(\sum_{n=1}^{N}p_n\frac{\delta^2_n}{\left(2^{B_n}-1\right)^2} - \epsilon\right)+\lambda_4(a_1-l^{\mathrm{c}})\\
	& \quad +\sum_{n=1}^{N}\lambda_{5,n}(1-B_n),
	\end{split}
\end{equation}
where $\boldsymbol{\lambda}=(\lambda_{1,1},...,\lambda_{2,1},...,\lambda_{5,N})\geq0$ ('$\geq$' denotes the component-wise inequality) is the LM vector and $\lambda_{1,n},\lambda_{2,n},\lambda_{3},\lambda_{4},\lambda_{5,n},\,\forall n \in \mathcal{N}$, are associated with the constraints $\mathrm{C}_i, \,i=1,...,5$, respectively. Following that, the dual function is given as
\begin{equation}
	\mathcal{G}(\boldsymbol{\lambda})=\underset{l^{\mathrm{c}},\boldsymbol{E},\boldsymbol{B},\boldsymbol{l}^\mathrm{up}}{\text{min}}\mathcal{L}(l^{\mathrm{c}},\boldsymbol{l}^\mathrm{up},\boldsymbol{E},\boldsymbol{B},\boldsymbol{\lambda}),
\end{equation}
while the corresponding dual problem can be written as
\begin{equation}
	\underset{\boldsymbol{\lambda}\,}{\text{max \,}} \mathcal{G}(\boldsymbol{\lambda}).
\end{equation}
Since the primal problem is convex and the Slater's conditions are satisfied, strong duality holds, i.e., solving the dual is equivalent to solving the primal problem \cite{boyd}.
According to the Karush-Kuhn-Tucker (KKT) conditions, the optimal solution to the problem should satisfy
\begin{equation}
	\nabla\mathcal{L}(l^{\mathrm{c*}},\boldsymbol{l}^{\mathrm{up}*},\boldsymbol{E}^*,\boldsymbol{B}^*,\boldsymbol{\lambda}^*)=0.
\end{equation}
Thus, by taking $\frac{\partial\mathcal{L} }{\partial l^{\mathrm{c}}}=0$ and $\frac{\partial\mathcal{L} }{\partial E_n}=0,\, \forall n \in \mathcal{N}$, leads to
\begin{equation}
	l^{\mathrm{c*}}=\sqrt[3]{\frac{2\zeta\tau^3\sum_{n=1}^{N}\lambda^*_{2,n}c^3_nD^3_n}{1-\lambda^*_4}},
	\label{lc_optimal}
\end{equation}
and
\begin{equation}\label{E_n}
	E^*_n = l^{\mathrm{up*}}_nW\left(\frac{\lambda^*_{1,n}}{\lambda^*_{2,n}\ln(2)}-\frac{N_0}{g_n}\right), \, \forall n \in \mathcal{N}.
\end{equation}
From \eqref{E_n}, we observe that $\lambda^*_{2,n}\neq0, \, \forall n \in \mathcal{N}$. Taking this into account, according to the complementary slackness conditions which require
\begin{equation}
	\lambda^*_{2,n}\left(\frac{\zeta \tau^3c^3_nD^3_n}{{l^{\mathrm{c}*}}^2}+E^*_n - E^{\mathrm{max}}_n\right)=0, \quad \forall n \in \mathcal{N},
\end{equation}
 the constraint $\mathrm{C}_2$ should be satisfied with equality \cite{boyd}, leading to
\begin{equation}
	E^*_n =E^{\mathrm{max}}-\frac{\zeta \tau^3c^3_nD^3_n}{l^{\mathrm{c*}^2}}, \quad \forall n \in \mathcal{N}.
	\label{e_optimal}
\end{equation}
This is reasonable, since it indicates that users should utilize their whole available energy, towards minimizing the objective function.
Following that, $\frac{\partial\mathcal{L} }{\partial l^{\mathrm{up}}_n}=0, \forall n \in \mathcal{N}$, it holds that
\begin{equation}\label{l_n}
	\frac{W\lambda^*_{1,n}}{\ln(2)}\frac{-(1+\frac{b}{l^{\mathrm{up*}}_n})+1+(1+\frac{b}{l^{\mathrm{up*}}_n})\ln\left(1+\frac{b}{l^{\mathrm{up*}}_n}\right)}{1+\frac{b}{l^{\mathrm{up*}}_n}}=1,
\end{equation}
where $b=\frac{E^*_ng_n}{WN_0}$. The manipulation of \eqref{l_n}, results to 
\begin{equation}
	l^{\mathrm{up*}}_n=-\frac{g_nE^*_n}{WN_0(1+\mathcal{W}^{-1}_0(\psi_n))}, \quad \forall n \in \mathcal{N},
	\label{ln_optimal}
\end{equation}
where $\mathcal{W}_0$ is the principal branch of the Lambert W function \cite{lambert} and $\psi_n$ is given by
\begin{equation}
	\psi_n=-\frac{2^{-\frac{1}{W\lambda^*_{1,n}}}}{e}, \quad \forall n \in \mathcal{N}.
\end{equation}
Furthermore, it is easy to verify from \eqref{l_n} that $\lambda^*_{1,n}\neq0, \, \forall n \in \mathcal{N}$, since the case $\lambda^*_{1,n}=0$, leads to a contradiction. This result indicates that users should spend their resources so as to transmit exactly $S_n$ bits to the server, which is observed from the right-hand-side of $\mathrm{C}_1$. Thus, $\mathrm{C}_1$ is satisfied with equality, yielding
\begin{equation}
	B^*_n = \frac{l^{\mathrm{up*}}_nW}{d}\log_2\left(1+\frac{g_nE^*_n}{l^{\mathrm{up*}}_nWN_0}\right)-\frac{m}{d}-1, \quad \forall n \in \mathcal{N}.
	\label{B_optimal}
\end{equation}
Finally, by taking $\frac{\partial\mathcal{L} }{\partial B_n}=0,\, \forall n \in \mathcal{N}$, yields
\begin{equation}\label{B_n_LM}
\left(2^{B^*_n}-1\right)^3=\frac{2\lambda^*_3\ln(2)p_n\delta^2_n}{d\lambda^*_{1,n}-\lambda^*_{5,n}}2^{B^*_n}.
\end{equation}
From \eqref{B_n_LM}, it is obvious that $\lambda^*_3 \neq 0$, since the case $\lambda^*_3 = 0$ leads to $B^*_n=0$, which is infeasible. Therefore, also constraint $C_3$ is satisfied with equality.
\par
According to the previous analysis, the optimal variables $l^{\mathrm{c}*},\boldsymbol{E}^*,\boldsymbol{l}^\mathrm{up*},\boldsymbol{B}^*$ have been given in closed forms, in terms of the LMs, by the equations \eqref{lc_optimal},\eqref{e_optimal},\eqref{ln_optimal},\eqref{B_optimal}, respectively. Note that given the LMs, each optimization variable can be directly calculated, with the aforementioned order of appearance, by using the respective equations.  Subsequently, the LMs can be updated iteratively via a subgradient method, while the optimal variables can be calculated through the LMs. Note that the equations which have not been used for calculating the primal variables, can be utilized for updating the LMs.
\par
 Following this analysis, recall that $B_n, \,\forall n \in \mathcal{N}$, should finally take integer values.  Therefore, after obtaining the solutions to the problem, $B^*_n$ should be rounded to the smallest integer which is greater or equal to $B^*_n$. Thus we set $\tilde{B}^*_n=\lceil B^*_n \rceil$, in order to guarantee that $\mathrm{C}_3$ is still satisfied, i.e., the quantization error tolerance constraint is not violated. However, since with the selection of $B^*_n$ it was previously shown that $\mathrm{C}_1$ is satisfied with equality, by plugging $\tilde{B}^*_n$ into the problem, $\mathrm{C}_1$ will be now violated, due to the fact that $\tilde{B}^*_n>B^*_n$. Therefore, to address this issue,  the problem in \eqref{primary2} has to be solved once again for a fixed value of $\tilde{B}^*_n$, i.e.,
\begin{equation}
\begin{aligned}
& &  &\underset{l^{\mathrm{c}},\boldsymbol{E},\boldsymbol{l}^\mathrm{up};\tilde{\boldsymbol{B}}^*}{\text{min}} \quad
\text{\emph{{$l^{\mathrm{c}}+\sum_{n=1}^{N}l^{\mathrm{up}}_n$}}} \\ 
& & &\textbf{\emph{s.t.}}
\quad \mathrm{C}_{1}, \mathrm{C}_{2}, \mathrm{C}_{4}
\end{aligned}\label{resolve}
\end{equation}
and finally the optimal variables can be obtained. It should be clarified that \eqref{resolve} can be solved similarly to \eqref{primary2} for a fixed $\tilde{\boldsymbol{B}}^*$. To this end, $\boldsymbol{f}^*$ is given by \eqref{fn_optimal}, which concludes the overall solution to the problem.
\section{Numerical Experiments and Performance Evaluation}
For the simulation results, we assume that $N=10$ users are uniformly distributed in a circle with radius 1000m, while the server/BS is located at the center of the circle. The rest of the simulation parameters are presented in Table \ref{parameters}, and retain their respective values unless specified otherwise.
\begin{table}
	\caption{Simulation Parameters}
		\label{parameters}
	\centering
	\begin{tabular}{ P{1.5cm}|P{1.5cm}||P{1.5cm}|P{2cm}}
		\hline
		\centering
		\textbf{Parameter} & \centering \textbf{Value} & \textbf{Parameter} & \textbf{Value}\\
		\hline\hline
		$f^{\mathrm{max}}_n$ & $1.5\,\mathrm{GHz}$ & $D_n$ & 1 Mbit \\
		\hline
		$W$ & 0.3 MHz & $N_0$ & -174 dBm/Hz \\
		\hline
		$\zeta$ & $10^{-27}$ & $c_n$ & $\sim \mathcal{U}(10,40)$ \\
		\hline
		$N$ & 10 users & $d_n$ & $\sim \mathcal{U}(0,1000\mathrm{m})$ \\
		\hline
		$E_n^\mathrm{max}$ & 0.3 Joule & m &  64 bits \\
		\hline
		$d$ & 23820 & $\beta$ &  3.75 \\
		\hline
	\end{tabular}
\end{table}
\par
We select the FL task to be the image classification on the widely-known MNIST dataset \cite{mnist}. We assume that each user carries 200 data samples and trains a fully-connected feed-forward neural network with a single hidden layer, consisting of 30 nodes, while the \textit{softmax} is utilized as an activation function at the output layer. Thus, the total model parameters are $d=23860$ (i.e., $784\times30+30\times10=23820$ weights and $30+10=40$ biases). The mini-batch size has been set as $\vert \boldsymbol{\xi}^{i}_n(t)\vert=50, \, \forall n,i,t.$ Moreover, the Adam optimizer is utilized for the local training \cite{adam}. Following that, we consider two different cases of training data distributions. Firstly, for the case of IID data distribution among users, the training data is shuffled and randomly assigned to each user. Secondly, for the non-IID scenario, the training data is sorted by labels and each user is equipped only with 5 labels. For both cases, the datasets among users are non-overlapping.
\subsection{Effects of the quantization error tolerance on the convergence time and model accuracy}
\begin{figure}[t!]
	\centering
	\includegraphics[width=0.9\linewidth]{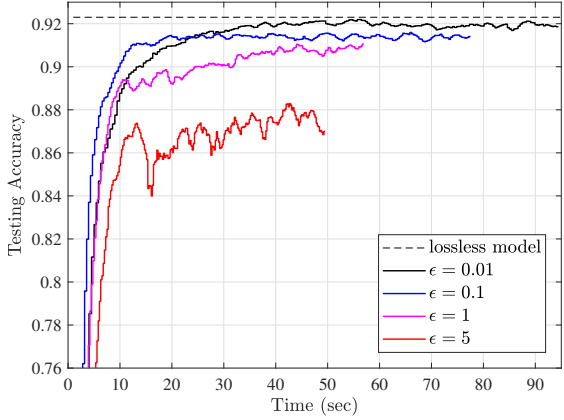}
	\caption{Testing accuracy versus time, for the IID scenario.}
	\label{acc1}
\end{figure}

\begin{figure}[t!]
	\centering
	\includegraphics[width=0.9\linewidth]{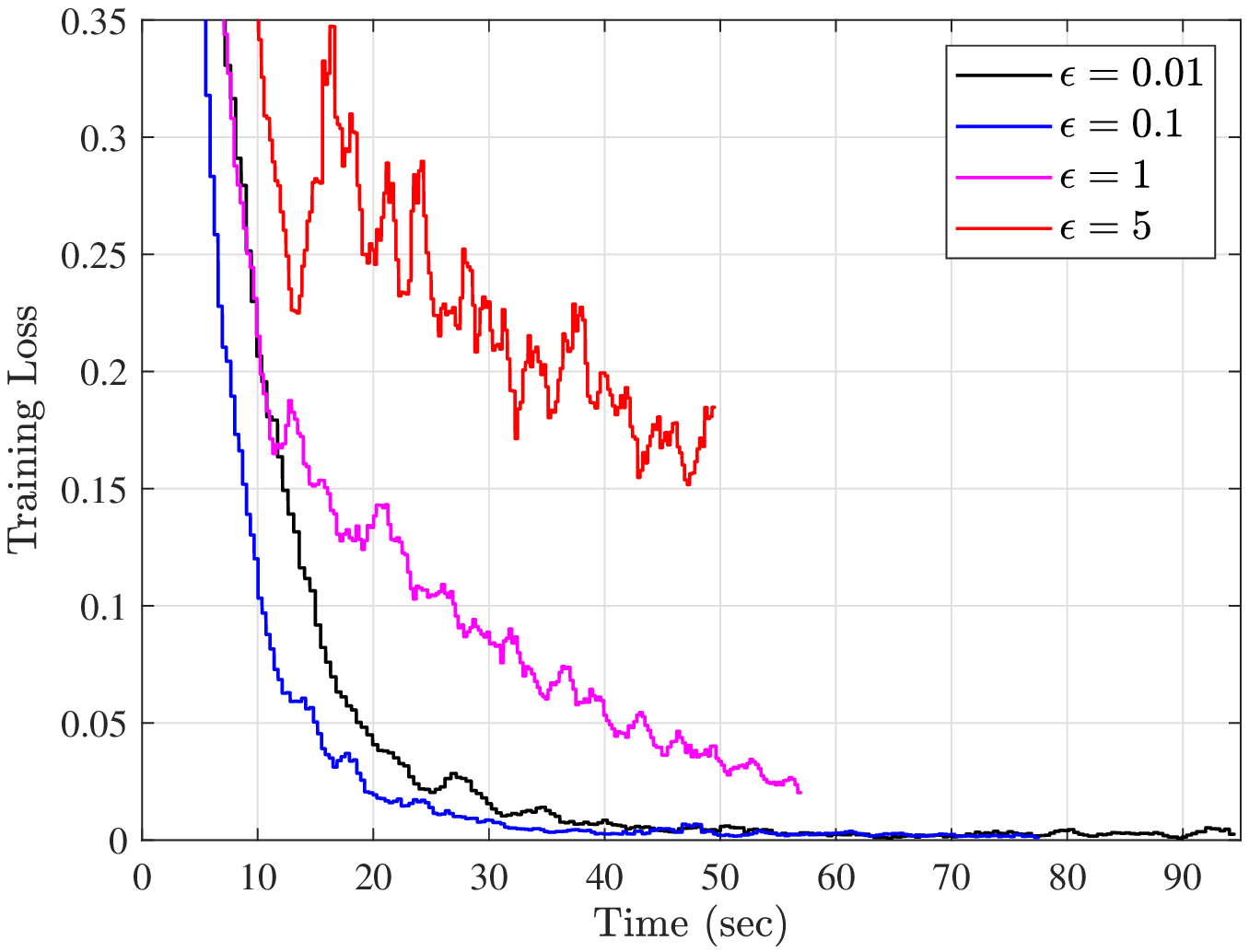}
	\caption{Training loss versus time, for the IID scenario.}
	\label{loss1}
\end{figure}

\begin{figure}[t!]
	\centering
	\includegraphics[width=0.9\linewidth]{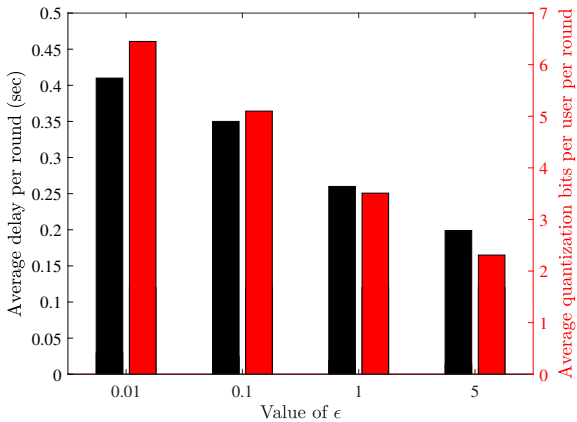}
	\caption{Average delay per round and average quantization bits per user per round, for the IID scenario.}
	\label{barplot}
\end{figure}

In Fig. \ref{acc1} and Fig. \ref{loss1}, the testing accuracy and training loss, respectively, are evaluated for various values of the quantization error tolerance, $\epsilon$. Note that the proposed optimization method has been utilized in order to extract all figures. In the considered simulations, we set $\tau=2$ local iterations for each user, while we set the total number of global communication rounds equal to $T=225$. It should be highlighted though, that the x axis illustrates the time in seconds, and not purely the evolution of the global rounds. We made this choice, since  for different tolerance values, the duration of a global round also varies. In this manner, the comparison on the convergence time between various values of $\epsilon$ can be fairly conducted. Also, we clarify that total the training time in seconds, given the $T$ global rounds, differs among different choices of $\epsilon$. From Fig. \ref{acc1}, it is evident that for smaller $\epsilon$, higher testing accuracy is achieved, which also approximates the performance of the lossless model, i.e., the FL model without applying quantization. Moreover, high values of $\epsilon$ may lead to decreased model accuracy, e.g., when $\epsilon=5$. Another interesting observation comes as follow. The case $\epsilon=0.1$, demonstrates the highest convergence rate in the early stages of the training process, which is also greater than the case $\epsilon=0.01$. This outcome is related with the duration of each global communication round. Smaller values of $\epsilon$, translate to the selection of more quantization bits, which in turn result in increased transmission latency during each communication round. More specifically, in Fig. \ref{barplot}, the average delay per global round and the average number of quantization bits per user per  global round are presented. It can been seen that the case $\epsilon=0.01$ presents the highest average delay per round, among the rest choices of $\epsilon$. Therefore, this fact can slow down the convergence speed in the early stages of the training, although finally the highest accuracy is achieved, owing this to the selection of more quantization bits which guaranty high precision. 
\par 
At this point, it is significant to highlight the following observation.  The cases $\epsilon=1$ and $\epsilon=5$ do not really contribute neither in model accuracy nor in fast convergence, since they are totally outperformed by the cases  $\epsilon=0.1$ and $\epsilon=0.01$, in both aspects. Even when targeting smaller accuracy values, the higher-$\epsilon$ cases fail to converge faster than the cases $\epsilon=0.1$ and $\epsilon=0.01$. An interpretation of this result is the following: The large number of communication rounds until convergence, which occur from the low precision quantization, prevails over the low-latency per round. Therefore, by selecting a relatively loose quantization error tolerance, which is equivalent to utilizing a few quantization bits, may not result in any gains or offer any benefits. This contradicts the fact which implies that by using a small number of quantization bits, communication efficiency is always achieved. 
\par
\begin{figure}[h!]
	\centering
	\includegraphics[width=0.9\linewidth]{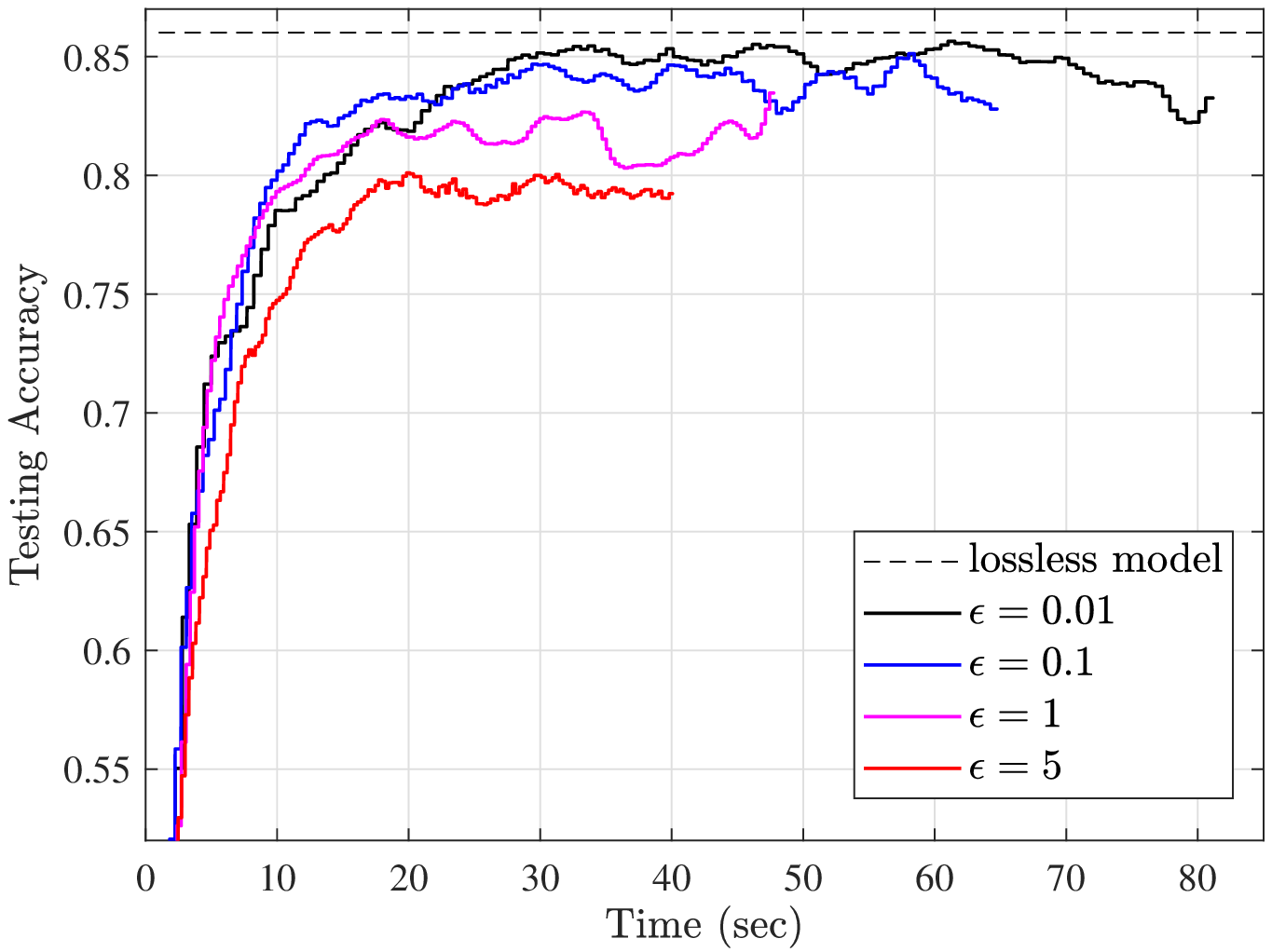}
	\caption{Testing accuracy versus time, for the NON-IID scenario.}
	\label{acc2}
\end{figure}
\begin{figure}[h!]
	\centering
	\includegraphics[width=0.9\linewidth]{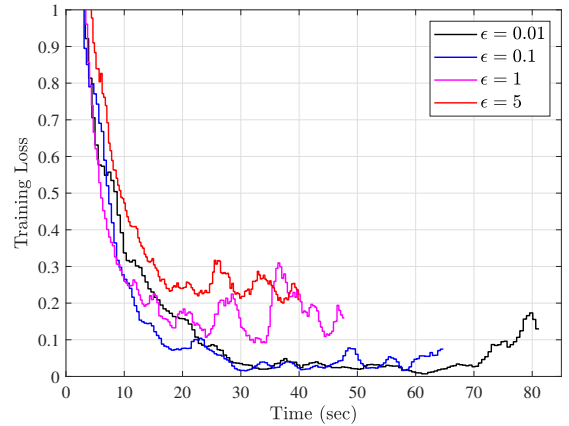}
	\caption{Training loss versus time, for the NON-IID scenario.}
	\label{loss2}
\end{figure}

Accordingly, in Fig. \ref{acc2} and  Fig. \ref{loss2}, the testing accuracy and training loss are evaluated for the NON-IID scenario. For this example, we set $\tau=3$ and $T=150$. Similar behavior in comparison with the IID case is observed, while the model accuracy is degraded.

\subsection{Comparison with baseline schemes}
\begin{figure}[h!]
	\centering
	\includegraphics[width=0.9\linewidth]{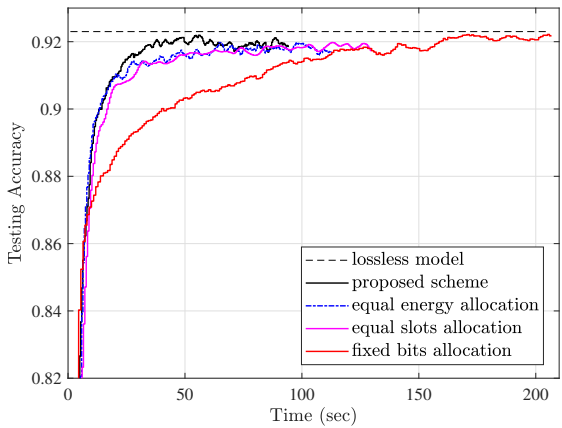}
	\caption{Comparison of the proposed scheme with baseline schemes.}
	\label{acc_fixed}
\end{figure}

 In Fig. \ref{acc_fixed}, we compare the performance of the proposed optimization scheme, with some baseline schemes. Firstly, we consider the \textit{fixed bits allocation} baseline scheme, where the same optimization method is adopted, but the number of quantization bits is pre-assigned as $\tilde{B}_n=16, \, \forall n \in \mathcal{N}$, in each global round. Moreover, the \textit{equal slots allocation} scheme also exploits the proposed optimizations' method, however it assigns equal duration uplink time slots to all users. Finally, the \textit{equal energy allocation} scheme assigns equal energy values for both the computation and transmission phase, while the rest of the optimization is conducted according to the proposed method. For the considered simulation, we set the quantization error tolerance $\epsilon=0.01$, consider an IID scenario and set $T=225$ global rounds.  By observing Fig. \ref{acc_fixed}, it can be seen that all schemes demonstrate almost identical testing accuracy, which is related with the selected quantization error tolerance. However, it is clearly seen that the proposed scheme dominates all baseline schemes, in terms of convergence time, which is the objective goal of the proposed optimization. Thus, Fig. \ref{acc_fixed} highlights the significance of the proposed scheme, which jointly takes into account the communication and computation resources, as well as the quantization bits allocation.  Also, the case of \textit{fixed bits allocation} leads to very high latency until convergence, without offering further accuracy gain. Therefore, it is evident that when aiming towards fast convergence of the FL process,  the number of quantization bits ought to be wisely selected.
\subsection{The dynamical adjustment of quantization error tolerance}
\begin{figure}[h!]
	\centering
	\includegraphics[width=0.9\linewidth]{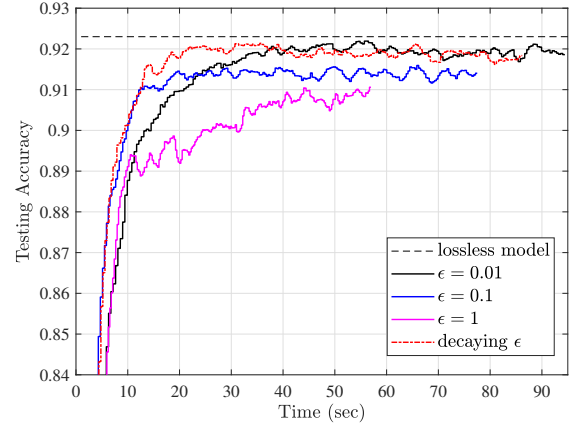}
	\caption{The impact of dynamically adjusting $\epsilon$, on the convergence rate and testing accuracy.}
	\label{d_e}
\end{figure}

In the continue, we examine the effects of dynamically adjusting the quantization error tolerance $\epsilon$ throughout the training process. Recall that in Theorem 1, it was evident that in the early training stages, the quantization error has not large impact in the  optimality gap. Driven by this fact, we now focus on decaying $\epsilon$ along with the evolution of the global rounds, i.e., allowing higher values of tolerance in the early stages of the training and gradually decreasing it. Specifically, we consider that the relation between the error tolerance in the $(i+1)$-th and $i$-th global round, is given as
\begin{equation}
	\epsilon(i+1)=r\cdot\epsilon(i),  
\end{equation}
where $0<r<1$ is a constant, $i=1,...,223$, while we initialize $\epsilon(0)=0.1$. By setting $r=10^{-1/224}$, it is easy to verify that $\epsilon(224)=0.01$, i.e., $\epsilon$ is equal to 0.1 in the first round and equal to 0.01 in the final round, given that $T=225$. Following that, in Fig. \ref{d_e}, we compare the performance of the considered technique, i.e., \textit{decaying $\epsilon$}, with the standard cases of a constant $\epsilon$ throughout the training. It can be observed that by dynamically decreasing $\epsilon$, the convergence rate is significantly increased. More specifically, in the very early stages of the training, higher $\epsilon$ values contribute to fast communication with the server, while along with the reduction of $\epsilon$, the precision is gradually increased. This policy results in fast convergence and notable performance, in comparison with the rest fixed $\epsilon$ cases and especially with the stringent case, where $\epsilon=0.01$. Specifically, although the testing accuracy is almost identical between the \textit{decaying $\epsilon$} policy and the case $\epsilon = 0.01$, the former converges after 22 seconds of training, while the latter after about 40 seconds. Therefore, the effectiveness of decreasing the quantization error tolerance  along with the evolution of the training, which is equivalent to gradually increasing the number of quantization bits throughout the training, is corroborated.


\section{Conclusions}
In this paper, we studied and optimized the performance of FL over wireless networks by considering the quantization of the local model parameters. More specifically, we have jointly optimized the communication and computation resources, as well as the quantization bits allocation, focusing on minimizing the total convergence time of FL subject to energy constraints and quantization error tolerance. The optimization problem was coupled with the convergence analysis, aiming to control the impact of the quantization error and subsequently balance the trade-off between model accuracy and fast convergence. Simulations are conducted, where the considered trade-off is examined and the effectiveness of the proposed method in accelerating the convergence speed, is verified. Also, the results indicate that the selection of the quantization error tolerance is critical for achieving enhanced performance in FL, while efficient techniques are presented which result in increased convergence rate.

\section*{Appendix A \\ Proof of Theorem 1}
For proving Theorem 1, we adopt the methodologies of \cite{li2019} and \cite{amiri2021}.
Firstly, we define the auxiliary variable $\boldsymbol{v}(t)$ as 
\begin{equation}
	\boldsymbol{v}(t+1)=\boldsymbol{w}(t)+\sum_{n=1}^{N}p_n\Delta \boldsymbol{w}_n(t),
\end{equation}
which represents a lossless model's update during the $(t+1)$-th round.
Recall that 
\begin{equation}
\boldsymbol{w}(t+1)=\boldsymbol{w}(t)+\sum_{n=1}^{N}p_nQ(\Delta \boldsymbol{w}_n(t)).
\end{equation}
Following that, we have
\begin{equation}\label{w_t+1 - w*}
	\begin{split}
	&\norm{\boldsymbol{w}(t+1)- \boldsymbol{w}^*}^2  \\
	&\qquad = \norm{\boldsymbol{w}(t+1)-\boldsymbol{v}(t+1)+\boldsymbol{v}(t+1)-\boldsymbol{w}^*}^2\\ & \qquad = \norm{\boldsymbol{w}(t+1)-\boldsymbol{v}(t+1)}^2+\norm{\boldsymbol{v}(t+1)-\boldsymbol{w}^*}^2\\
	&\qquad \quad + 2\langle \boldsymbol{w}(t+1)-\boldsymbol{v}(t+1),\boldsymbol{v}(t+1)-\boldsymbol{w}^*\rangle.
	\end{split}
\end{equation}
In the continue, the average of the right-hand-side terms in \eqref{w_t+1 - w*} are bounded and presented in Lemmas 2-4 respectively.
\begin{lemma}
	We have
	\begin{equation*}
		\mathbb{E}\left[\norm{\boldsymbol{w}(t+1)-\boldsymbol{v}(t+1)}^2\right] \leq \sum_{n=1}^{N}p_nJ^2_n(t).
	\end{equation*}
\end{lemma}
\begin{IEEEproof}
\begin{equation}
\begin{split}
	&\mathbb{E}\left[\norm{\boldsymbol{w}(t+1)-\boldsymbol{v}(t+1) }^2\right]\\ 
	& \qquad =\mathbb{E}\left[\norm{\sum_{n=1}^{N}p_n(Q(\Delta \boldsymbol{w}_n(t))-\Delta \boldsymbol{w}_n(t))}^2\right]\\
	 & \qquad \overset{(a)}{\leq} \sum_{n=1}^{N}p_n\mathbb{E}\left[\norm{(Q(\Delta \boldsymbol{w}_n(t))-\Delta \boldsymbol{w}_n(t))}^2\right]\\
	 & \qquad \overset{(b)}{\leq} \sum_{n=1}^{N}p_nJ^2_n(t),
\end{split}
\end{equation}
where (a) follows from the convexity of $\norm{\cdot}^2$ and the fact that $\sum_{n=1}^{N}p_n=1$, while (b) follows from Lemma 1.
\end{IEEEproof}
\begin{lemma}
	We have
	\begin{equation}
 \mathbb{E}\left[\norm{\boldsymbol{v}(t+1)-\boldsymbol{w}^*}^2\right]
		 \leq  -\mu\eta(t)\mathbb{E}\left[\norm{\boldsymbol{w}(t)-\boldsymbol{w}^*}^2\right]+\eta^2(t)U
	\end{equation}
	where 
	\begin{equation}
	U=\tau^2\sum_{n=1}^{N}\sigma^2_n+\tau G^2+ 2L\tau^2\Gamma
	+ (\mu+2)\frac{\tau(\tau-1)(2\tau-1)}{6}G^2.
	\end{equation}
\end{lemma}
\begin{IEEEproof}
See Appendix B.
\end{IEEEproof}
\begin{lemma}
	We have
	\begin{equation}\label{lemma4}
		\mathbb{E}\left[2\langle \boldsymbol{w}(t+1)-\boldsymbol{v}(t+1),\boldsymbol{v}(t+1)-\boldsymbol{w}^*\rangle\right]=0.
	\end{equation}
\end{lemma}
\begin{IEEEproof}
	Since it holds
	\begin{equation}
			\mathbb{E}\left[Q(\Delta \boldsymbol{w}_n(t))\right]=\Delta \boldsymbol{w}_n(t),
	\end{equation}
	and 
	\begin{equation}
		\boldsymbol{w}(t+1)-\boldsymbol{v}(t+1)=\sum_{n=1}^{N}p_n(Q(\Delta \boldsymbol{w}_n(t))-\Delta \boldsymbol{w}_n(t)),
	\end{equation}
it is straightforward to conclude to \eqref{lemma4}.
\end{IEEEproof}
\par
Following that, by combining the results in Lemmas 2-4, \eqref{w_t+1 - w*} leads to
\begin{equation}\label{basic}
\begin{split}
\mathbb{E}\left[\norm{\boldsymbol{w}(t+1)-\boldsymbol{w}^*}^2\right]&
\leq(1-\eta(t)\mu)\mathbb{E}\left[\norm{\boldsymbol{w}(t)-\boldsymbol{w}^*}^2\right]\\
& \quad + \eta^2(t)U + \sum_{n=1}^{N}p_nJ^2_n(t).
\end{split}
\end{equation}
Let $\Delta_t=\mathbb{E}\left[\norm{\boldsymbol{w}(t)-\boldsymbol{w}^*}^2\right]$.  \eqref{basic} can be re-written as
\begin{equation}
	\Delta_{t+1}\leq (1-\eta(t)\mu)\Delta_t + \eta^2(t)U + \sum_{n=1}^{N}p_nJ^2_n(t).
\end{equation}
Next, we will show that $\Delta_t \leq \frac{\nu}{\gamma + t}+\Psi(t)$ where
\begin{equation}
\Psi(x)\triangleq \sum_{j=0}^{x-1}\sum_{n=1}^{N}p_nJ^2_n(j)\prod_{i=j+1}^{x-1}(1-\eta(i)\mu), \quad x \geq 1,
\end{equation}
by selecting a diminishing learning rate $\eta(t)=\frac{\beta}{\gamma+t}$, with $\beta\geq\frac{1}{\mu}, \, \gamma\geq\beta$ such that $\eta(0)\leq1$, $\gamma\geq\beta\mu$ such that $\eta(0)\leq\frac{1}{\mu}$ and $\nu \leq \mathrm{max}\left\{\frac{\beta^2U}{\beta\mu-1},\gamma\Delta_0\right\}$. Also, since we required $\eta(t)\leq\frac{1}{L\tau}$, it should also hold $\gamma\geq\frac{L}{\mu}$. Similarly to \cite{li2019}, via induction we have
\begin{equation}\label{Delta_t+1}
\begin{split}
\Delta_{t+1}&\leq (1-\eta(t)\mu)\Delta_t + \eta^2(t)U + \sum_{n=1}^{N}p_nJ^2_n(t)\\
& \leq \left(1-\frac{\beta\mu}{\gamma+t}\right)\left(\frac{\nu}{\gamma+t}+\Psi(t)\right)+\frac{\beta^2U^2}{(\gamma+t)^2}\\
&=\frac{t+\gamma-1}{(t+\gamma)^2}\nu+\left(\frac{\beta^2U^2}{(\gamma+t)^2}-\frac{\beta\mu-1}{(t+\gamma)^2}\nu\right)\\
& \quad+\Psi(t+1)\\
&\leq \frac{\nu}{t+\gamma+1}+\Psi(t+1).
\end{split}
\end{equation}
Following that, we have
\begin{equation}\label{v}
	\begin{split}
	\nu &\leq \mathrm{max}\left\{\frac{\beta^2U}{\beta\mu-1},\gamma\Delta_0\right\}\leq \frac{\beta^2U}{\beta\mu-1}+\gamma\Delta_0\\
	& \overset{\beta=\frac{2}{\mu}}{\leq} \frac{4U}{\mu^2}+\gamma\mathbb{E}\left[\norm{\boldsymbol{w}(0)-\boldsymbol{w}^*}^2\right].\\
	\end{split}
\end{equation}
By substituting \eqref{v} in \eqref{Delta_t+1} and by using the fact that $F(\cdot)$ is L-smooth, which gives
\begin{equation}
\begin{split}
	F(\boldsymbol{w}(T))-F(\boldsymbol{w}^*)&\leq\langle \boldsymbol{w}(T)-\boldsymbol{w}^*,\nabla F(\boldsymbol{w}^*)\rangle\\
	&\quad +\frac{L}{2}\norm{\boldsymbol{w}(T)-\boldsymbol{w}^*}^2\\
	&\leq\frac{L}{2}\norm{\boldsymbol{w}(T)-\boldsymbol{w}^*}^2,
\end{split}
\end{equation}
since $\nabla F(\boldsymbol{w}^*)=0$,
 the proof of Theorem 1 is completed.
\section*{Appendix B \\ Proof of Lemma 3}
Firstly, we have
\begin{equation}\label{basic_on_proof}
\begin{split}
&\mathbb{E}\left[\norm{\boldsymbol{v}(t+1)-\boldsymbol{w}^*}^2\right] \\
& \qquad = \mathbb{E}\left[\norm{\boldsymbol{w}(t)+\sum_{n=1}^{N}p_n\Delta \boldsymbol{w}_n(t)-\boldsymbol{w}^*}^2\right]\\
& \qquad = \mathbb{E}\left[\norm{\boldsymbol{w}(t)-\boldsymbol{w}^*}^2\right]+\underbrace{\mathbb{E}\left[\norm{\sum_{n=1}^{N}p_n\Delta \boldsymbol{w}_n(t)}^2\right]}_{A_1}\\
& \qquad \quad + \underbrace{2\mathbb{E}\left[\Big\langle\boldsymbol{w}(t)-\boldsymbol{w}^*,\sum_{n=1}^{N}p_n\Delta \boldsymbol{w}_n(t)\Big\rangle\right]}_{A_2}.
\end{split}
\end{equation}
For $A_1$, we have that
\begin{equation}\label{A1}
\begin{split}
A_1 & = \mathbb{E}\left[\norm{\sum_{n=1}^{N}p_n\left(-\eta(t)\sum_{i=1}^{\tau}\nabla F_n(\boldsymbol{w}^{i-1}_n(t),\boldsymbol{\xi}^{i-1}_n(t))\right)}^2\right]\\
& \overset{(a)}{\leq} \eta^2(t)\tau\sum_{n=1}^{N}p_n\sum_{i=1}^{\tau}\mathbb{E}\left[\norm{\nabla F_n(\boldsymbol{w}^{i-1}_n(t),\boldsymbol{\xi}^{i-1}_n(t))}^2\right]\\
& = \eta^2(t)\tau\sum_{n=1}^{N}p_n\sum_{i=1}^{\tau}\mathbb{E}\Big[\Vert\nabla F_n(\boldsymbol{w}^{i-1}_n(t),\boldsymbol{\xi}^{i-1}_n(t))\\
& \quad -\nabla F_n(\boldsymbol{w}^{i-1}_n(t))+\nabla F_n(\boldsymbol{w}^{i-1}_n(t))\Vert^2_2\Big]\\
& \overset{(b)}{\leq} \eta^2(t)\tau\left(\tau\sum_{n=1}^{N}\sigma^2_n+\sum_{n=1}^{N}p_n\sum_{i=1}^{\tau}\mathbb{E}\left[\norm{\nabla F_n(\boldsymbol{w}^{i-1}_n(t))}^2\right]\right)
\\
& \overset{(c)}{\leq} \eta^2(t)\tau^2\sum_{n=1}^{N}\sigma^2_n\\
& \quad + 2L\eta^2(t)\tau\sum_{n=1}^{N}p_n\sum_{i=1}^{\tau}\mathbb{E}\left[F_n(\boldsymbol{w}^{i-1}_n(t))-F^*_n\right],
\end{split}
\end{equation}
where (a) follows from the convexity of $\norm{\cdot}^2$, (b) from Assumption 4 and $\mathbb{E}[\nabla F_n(\boldsymbol{w}^{i-1}_n(t),\boldsymbol{\xi}^{i-1}_n(t))]=\nabla F_n(\boldsymbol{w}^{i-1}_n(t))$, while (c) from the L-smoothness of $F_n$, which implies that \cite{boyd}:
\begin{equation}\label{boyd_eq}
\norm{\nabla F_n(\boldsymbol{w}^{i-1}_n(t))}^2 \leq 2L(F_n(\boldsymbol{w}^{i-1}_n(t))-F^*_n).
\end{equation}
In the following we bound the last term in \eqref{basic_on_proof}, $A_2$, as:
\begin{equation}\label{A2}
\begin{split}
& A_2  = 2\sum_{n=1}^{N}p_n\mathbb{E}\left[\langle\boldsymbol{w}(t)-\boldsymbol{w}^*,\Delta \boldsymbol{w}_n(t)\rangle\right]\\
& = 2\eta(t)\sum_{n=1}^{N}p_n\sum_{i=1}^{\tau}\mathbb{E}\left[\langle \boldsymbol{w}^*-\boldsymbol{w}(t),\nabla F_n(\boldsymbol{w}^{i-1}_n(t),\boldsymbol{\xi}^{i-1}_n(t))\rangle\right]\\
& = 2\eta(t) \times \\ 
&\quad \Bigg[\underbrace{\sum_{n=1}^{N}p_n\sum_{i=1}^{\tau}\mathbb{E}\left[\langle \boldsymbol{w}^{i-1}_n(t)-\boldsymbol{w}(t),\nabla F_n(\boldsymbol{w}^{i-1}_n(t),\boldsymbol{\xi}^{i-1}_n(t))\rangle\right]}_{B_1}\\
& \, + \underbrace{\sum_{n=1}^{N}p_n\sum_{i=1}^{\tau}\mathbb{E}\left[\langle \boldsymbol{w}^*- \boldsymbol{w}^{i-1}_n(t),\nabla F_n(\boldsymbol{w}^{i-1}_n(t),\boldsymbol{\xi}^{i-1}_n(t))\rangle\right]}_{B_2}\Bigg].
\end{split}
\end{equation}
Next, for $B_1$ we have
\begin{equation}\label{B1}
\begin{split}
B_1 &  \overset{(a)}{\leq} \eta(t)\sum_{n=1}^{N}p_n\sum_{i=1}^{\tau}\mathbb{E}\Big[\frac{1}{\eta(t)}\norm{\boldsymbol{w}^{i-1}_n(t)-\boldsymbol{w}(t)}^2\\
& \quad +\eta(t)\norm{\nabla F_n(\boldsymbol{w}^{i-1}_n(t),\boldsymbol{\xi}^{i-1}_n(t))}^2\Big]\\
& \overset{(b)}{\leq} \sum_{n=1}^{N}p_n\sum_{i=1}^{\tau}\mathbb{E}\left[\norm{\boldsymbol{w}^{i-1}_n(t)-\boldsymbol{w}(t)}^2\right]+\eta^2(t)\tau G^2,
\end{split}
\end{equation}
where (a) follows from the Cauchy-Schwarz inequality in combination with the inequality
\begin{equation}\label{AM-GM}
2\left(\frac{x}{\sqrt{\eta(t)}}\right)(\sqrt{\eta(t)}y) \leq \frac{x^2}{\eta(t)}+\eta(t)y^2,
\end{equation}
while (b) follows from Assumption 3. Next, we bound $B_2$ from \eqref{A2} as:
\begin{equation}
\begin{split}
B_2 & \leq 2\eta(t)\sum_{n=1}^{N}p_n\sum_{i=1}^{\tau}\mathbb{E}\left[\langle \boldsymbol{w}^*- \boldsymbol{w}^{i-1}_n(t),\nabla F_n(\boldsymbol{w}^{i-1}_n(t))\rangle\right]\\
& \overset{(a)}{\leq} 2\eta(t)\sum_{n=1}^{N}p_n\sum_{i=1}^{\tau}\mathbb{E}\Big[F_n(\boldsymbol{w}^*)-F_n(\boldsymbol{w}^{i-1}_n(t))\\
& \quad-\frac{\mu}{2}\norm{\boldsymbol{w}^{i-1}_n(t)-\boldsymbol{w}^*}^2\Big],\\
\end{split}
\end{equation}
where (a) follows from the $\mu$-strong convexity of $F_n, \, \forall n \in \mathcal{N}$. By combining \eqref{A1} and \eqref{A2} we conclude to
\begin{equation}\label{A1+A2}
\begin{split}
A_1&+A_2 \\
& \leq \eta^2(t)\tau^2\sum_{n=1}^{N}\sigma^2_n + \sum_{n=1}^{N}p_n\sum_{i=1}^{\tau}\mathbb{E}\left[\norm{\boldsymbol{w}^{i-1}_n(t)-\boldsymbol{w}(t)}^2\right]\\
& \quad +\eta^2(t)\tau G^2-2\eta(t)\sum_{n=1}^{N}p_n\sum_{i=1}^{\tau}\mathbb{E}\left[\frac{\mu}{2}\norm{\boldsymbol{w}^{i-1}_n(t)-\boldsymbol{w}^*}^2\right]\\
& \quad + 2L\eta^2(t)\tau\sum_{n=1}^{N}p_n\sum_{i=1}^{\tau}\mathbb{E}\left[F_n(\boldsymbol{w}^{i-1}_n(t))-F^*_n\right]\\
&\quad - 2\eta(t)\sum_{n=1}^{N}p_n\sum_{i=1}^{\tau}\mathbb{E}\left[F_n(\boldsymbol{w}^{i-1}_n(t))-F_n(\boldsymbol{w}^*)\right]. \\
\end{split}
\end{equation}
Next, we bound the last two terms in \eqref{A1+A2}, which we denote as $C$. That gives
\begin{equation}
\begin{split}
C & = 2L\eta^2(t)\tau\sum_{n=1}^{N}p_n\sum_{i=1}^{\tau}\mathbb{E}\left[F_n(\boldsymbol{w}^{i-1}_n(t))-F^*_n\right]\\
&\quad - 2\eta(t)\sum_{n=1}^{N}p_n\sum_{i=1}^{\tau}\mathbb{E}\left[F_n(\boldsymbol{w}^{i-1}_n(t))-F_n(\boldsymbol{w}^*)\right] \\
& =- 2\eta(t)(1-L\eta(t)\tau)\sum_{n=1}^{N}p_n\sum_{i=1}^{\tau}\mathbb{E}\left[F_n(\boldsymbol{w}^{i-1}_n(t))-F^*_n\right]\\
& \quad +2\eta(t)\sum_{n=1}^{N}p_n\sum_{i=1}^{\tau}\mathbb{E}\left[F_n(\boldsymbol{w}^{*})-F^*_n\right].\\
\end{split}
\end{equation}
We can now write $C$ as
\begin{equation}\label{C_final}
	\begin{split}
		C & = - 2\eta(t)(1-L\eta(t)\tau)\\
		&\quad \times\sum_{n=1}^{N}p_n\sum_{i=1}^{\tau}\mathbb{E}\left[F_n(\boldsymbol{w}^{i-1}_n(t))-F(\boldsymbol{w}^*)\right]\\
		& \quad + \left(2\eta(t)-2\eta(t)(1-L\eta(t)\tau)\right)\\
		&\quad \times\sum_{n=1}^{N}p_n\sum_{i=1}^{\tau}\mathbb{E}\left[F(\boldsymbol{w}^*)-F^*_n\right]\\
		& \leq - 2\eta(t)(1-L\eta(t)\tau)\\
		&\quad \times\underbrace{\sum_{n=1}^{N}p_n\sum_{i=1}^{\tau}\mathbb{E}\left[F_n(\boldsymbol{w}^{i-1}_n(t))-F(\boldsymbol{w}^*)\right]}_{D}\\
		& \quad +2L\eta^2(t)\tau^2\Gamma.\\
	\end{split}
\end{equation}
To bound $D$ from \eqref{C_final}, we write
\begin{equation}
\begin{split}
D & \overset{(a)}{=} \sum_{n=1}^{N}p_n\sum_{i=1}^{\tau}\mathbb{E}\left[F_n(\boldsymbol{w}^{i-1}_n(t))-F(\boldsymbol{w}(t))\right]\\
& \quad + \sum_{n=1}^{N}p_n\sum_{i=1}^{\tau}\mathbb{E}\left[F_n(\boldsymbol{w}(t))-F(\boldsymbol{w}^*)\right]\\
& \geq \sum_{n=1}^{N}p_n\sum_{i=1}^{\tau}\mathbb{E}\left[\langle\nabla F_n(\boldsymbol{w}(t)),\boldsymbol{w}^{i-1}_n(t)-\boldsymbol{w}(t)\rangle\right]\\
& \quad+\tau\mathbb{E}\left[(F(\boldsymbol{w}(t))-F(\boldsymbol{w^*}))\right]\\
& \overset{(b)}{\geq} -\frac{1}{2}\sum_{n=1}^{N}p_n\sum_{i=1}^{\tau}\mathbb{E}\Big[\norm{\eta(t)\nabla F_n(\boldsymbol{w}(t))}^2\\
& +\frac{1}{\eta(t)}\norm{\boldsymbol{w}^{i-1}_n(t)-\boldsymbol{w}(t)}^2\Big]+\tau\mathbb{E}\left[F(\boldsymbol{w}(t))-F(\boldsymbol{w}^*)\right]\\
& \overset{(c)}{\geq} -\sum_{n=1}^{N}p_n\sum_{i=1}^{\tau}\mathbb{E}\Big[\eta(t)L(F_n(\boldsymbol{w}(t))-F(\boldsymbol{w^*}))\\
&+\frac{1}{2\eta(t)}\norm{\boldsymbol{w}^{i-1}_n(t)-\boldsymbol{w}(t)}^2\Big]+\tau\mathbb{E}\left[F(\boldsymbol{w}(t))-F(\boldsymbol{w}^*)\right],\\
\end{split}
\end{equation}
where in (a) we have used that $\sum_{n=1}^{N}p_nF_n(\boldsymbol{w}(t))=F(\boldsymbol{w}(t))$, (b) follows from Cauchy-Schwarz inequality and (c) from \eqref{boyd_eq}.
Next, by plugging $D$ in $C$ we get
\begin{equation}
\begin{split}
C & \leq  2\eta(t)(1-L\eta(t)\tau)\sum_{n=1}^{N}p_n\sum_{i=1}^{\tau}\mathbb{E}\Big[\eta(t)L(F_n(\boldsymbol{w}(t))\\
&\quad-F(\boldsymbol{w^*}))+\frac{1}{2\eta(t)}\norm{\boldsymbol{w}^{i-1}_n(t)-\boldsymbol{w}(t)}^2\Big] +2L\eta^2(t)\tau^2\Gamma\\
&\quad -2\eta(t)(1-L\eta(t)\tau)\mathbb{E}\left[\tau(F(\boldsymbol{w}(t))-F(\boldsymbol{w^*}))\right]\\
&\overset{(a)}{\leq} 2L\eta^2(t)\tau^2\Gamma + \sum_{n=1}^{N}p_n\sum_{i=1}^{\tau}\mathbb{E}\left[\norm{\boldsymbol{w}^{i-1}_n(t)-\boldsymbol{w}(t)}^2\right],\\
\end{split}
\end{equation}
where (a) holds for $\eta(t)\leq\frac{1}{\tau L}$, since $F(\boldsymbol{w}(t))-F(\boldsymbol{w^*}) \geq 0, \, \forall t$.

By plugging \eqref{A1} and \eqref{A2} in \eqref{A1+A2}, yields
\begin{equation}
\begin{split}
A_1&+A_2  \leq \eta^2(t)\tau^2\sum_{n=1}^{N}\sigma^2_n+\eta^2(t)\tau G^2+ 2L\eta^2(t)\tau^2\Gamma\\&\quad + 2\sum_{n=1}^{N}p_n\sum_{i=1}^{\tau}\mathbb{E}\left[\norm{\boldsymbol{w}^{i-1}_n(t)-\boldsymbol{w}(t)}^2\right]\\
& \quad - \mu\eta(t)\sum_{n=1}^{N}p_n\sum_{i=1}^{\tau}\mathbb{E}\left[\norm{\boldsymbol{w}^{i-1}_n(t)-\boldsymbol{w}^*}^2\right]\\
& \overset{(a)}{\leq} \eta^2(t)\tau^2\sum_{n=1}^{N}\sigma^2_n+\eta^2(t)\tau G^2+ 2L\eta^2(t)\tau^2\Gamma\\&\quad + 2\sum_{n=1}^{N}p_n\sum_{i=2}^{\tau}\mathbb{E}\left[\norm{\boldsymbol{w}^{i-1}_n(t)-\boldsymbol{w}(t)}^2\right]\\
& \quad - \mu\eta(t)\sum_{n=1}^{N}p_n\sum_{i=2}^{\tau}\mathbb{E}\left[\norm{\boldsymbol{w}^{i-1}_n(t)-\boldsymbol{w}^*}^2\right]\\
&\quad - \mu\eta(t)\mathbb{E}\left[\norm{\boldsymbol{w}(t)-\boldsymbol{w}^*}^2\right],\\
\end{split}
\end{equation}
and now we have
\begin{equation}\label{A1+A2_ineq}
	\begin{split}
		A_1&+A_2\\ &\overset{(b)}{\leq}\eta^2(t)\tau^2\sum_{n=1}^{N}\sigma^2_n+\eta^2(t)\tau G^2+ 2L\eta^2(t)\tau^2\Gamma\\
		&\quad + 2\sum_{n=1}^{N}p_n\sum_{i=2}^{\tau}\mathbb{E}\left[\norm{\boldsymbol{w}^{i-1}_n(t)-\boldsymbol{w}(t)}^2\right]\\
		& \quad+\mu\eta(t)(1-\eta(t))\Bigg(-(\tau-1)\mathbb{E}\left[\norm{\boldsymbol{w}(t)-\boldsymbol{w}^*}^2\right]\\
		&\quad +\frac{1}{\eta(t)}\sum_{n=1}^{N}p_n\sum_{i=2}^{\tau}\mathbb{E}\left[\norm{\boldsymbol{w}^{i-1}_n(t)-\boldsymbol{w}(t)}^2\right]\Bigg)\\
		&\quad - \mu\eta(t)\mathbb{E}\left[\norm{\boldsymbol{w}(t)-\boldsymbol{w}^*}^2\right],\\
	\end{split}
\end{equation}
where in (a) we used that $\boldsymbol{w}^0_n(t)\triangleq \boldsymbol{w}(t)$, while (b) follows from the fact that:
\begin{equation}
\begin{split}
-&\norm{\boldsymbol{w}^{i-1}_n(t)-\boldsymbol{w}^*}^2\\ 
&= -\norm{\boldsymbol{w}^{i-1}_n(t)-\boldsymbol{w}(t)}^2-\norm{\boldsymbol{w}(t)-\boldsymbol{w}^*}^2\\ & \quad -2\langle\boldsymbol{w}^{i-1}_n(t)-\boldsymbol{w}(t),\boldsymbol{w}(t)-\boldsymbol{w}^*\rangle\\
& \overset{(c)}{\leq} -\norm{\boldsymbol{w}^{i-1}_n(t)-\boldsymbol{w}(t)}^2-\norm{\boldsymbol{w}(t)-\boldsymbol{w}^*}^2\\
& \quad + \frac{1}{\eta(t)}\norm{\boldsymbol{w}^{i-1}_n(t)-\boldsymbol{w}(t)}^2+\eta(t)\norm{\boldsymbol{w}(t)-\boldsymbol{w}^*}^2\\
& = \left(\frac{1}{\eta(t)}-1\right)\norm{\boldsymbol{w}^{i-1}_n(t)-\boldsymbol{w}(t)}^2\\
& \quad-(1-\eta(t))\norm{\boldsymbol{w}(t)-\boldsymbol{w}^*}^2,
\end{split}
\end{equation}
where (c) follows from Cauchy-Schwarz inequality, compined with the inequality in \eqref{AM-GM}. By further expanding \eqref{A1+A2_ineq}, we get
\begin{equation}
	\begin{split}
	 A_1&+A_2\\
	 & \leq \eta^2(t)\tau^2\sum_{n=1}^{N}\sigma^2_n+\eta^2(t)\tau G^2+ 2L\eta^2(t)\tau^2\Gamma\\
	 & \quad -\mu\eta(t)(\tau-\eta(t)(\tau-1))\mathbb{E}\left[\norm{\boldsymbol{w}(t)-\boldsymbol{w}^*}^2\right]\\
	 & \quad + (2+\mu-\mu\eta(t))\sum_{n=1}^{N}p_n\sum_{i=2}^{\tau}\mathbb{E}\left[\norm{\boldsymbol{w}^{i-1}_n(t)-\boldsymbol{w}(t)}^2\right],\\
	 &\overset{(a)}{\leq} \eta^2(t)\tau^2\sum_{n=1}^{N}\sigma^2_n+\eta^2(t)\tau G^2+ 2L\eta^2(t)\tau^2\Gamma\\
	 & \quad -\mu\eta(t)\mathbb{E}\left[\norm{\boldsymbol{w}(t)-\boldsymbol{w}^*}^2\right]\\
	 & \quad + (2+\mu)\sum_{n=1}^{N}p_n\sum_{i=2}^{\tau}\mathbb{E}\left[\norm{\boldsymbol{w}^{i-1}_n(t)-\boldsymbol{w}(t)}^2\right],\\
	\end{split}\label{A1+A2_FINAL}
\end{equation}
where in (a) we have used that $0<\eta(t)\leq 1$, which also implies that $\tau-\eta(t)(\tau-1)\geq1$. 
Finally,
the last term in \eqref{A1+A2_FINAL}, can be bounded as follows:
\begin{equation}\label{last_in_3}
\begin{split}
&\sum_{n=1}^{N}p_n  \sum_{i=2}^{\tau} \mathbb{E}\left[\norm{\boldsymbol{w}^{i-1}_n(t)-\boldsymbol{w}(t)}^2\right]\\
& \qquad = \eta^2(t)\sum_{n=1}^{N}p_n\sum_{i=2}^{\tau}\mathbb{E}\left[\norm{\sum_{j=1}^{i}\nabla F_n(\boldsymbol{w}^{j-1}_n(t),\boldsymbol{\xi}^{j-1}_n(t))}^2\right]\\
&\qquad \overset{(a)}{\leq}  \eta^2(t)\sum_{i=2}^{\tau}i^2G^2
=
\eta^2(t)\frac{\tau(\tau-1)(2\tau-1)}{6}G^2,
\end{split}
\end{equation}
where (a) follows from the convexity of $\norm{\cdot}^2$ and  Assumption 3. By substituting \eqref{last_in_3} in \eqref{A1+A2_FINAL}, the proof of Lemma 3 is completed. 

\bibliographystyle{IEEEtran}
\bibliography{QuantizedFL}

\begin{thebibliography}{10}
\providecommand{\url}[1]{#1}
\csname url@samestyle\endcsname
\providecommand{\newblock}{\relax}
\providecommand{\bibinfo}[2]{#2}
\providecommand{\BIBentrySTDinterwordspacing}{\spaceskip=0pt\relax}
\providecommand{\BIBentryALTinterwordstretchfactor}{4}
\providecommand{\BIBentryALTinterwordspacing}{\spaceskip=\fontdimen2\font plus
\BIBentryALTinterwordstretchfactor\fontdimen3\font minus
  \fontdimen4\font\relax}
\providecommand{\BIBforeignlanguage}[2]{{%
\expandafter\ifx\csname l@#1\endcsname\relax
\typeout{** WARNING: IEEEtran.bst: No hyphenation pattern has been}%
\typeout{** loaded for the language `#1'. Using the pattern for}%
\typeout{** the default language instead.}%
\else
\language=\csname l@#1\endcsname
\fi
#2}}
\providecommand{\BIBdecl}{\relax}
\BIBdecl

\bibitem{letaief}
K.~B. {Letaief}, W.~{Chen}, Y.~{Shi}, J.~{Zhang}, and Y.~A. {Zhang}, ``The
  roadmap to {6G}: {AI} empowered wireless networks,'' \emph{IEEE Commun.
  Mag.}, vol.~57, no.~8, pp. 84--90, 2019.

\bibitem{sun2019}
Y.~Sun, M.~Peng, Y.~Zhou, Y.~Huang, and S.~Mao, ``Application of machine
  learning in wireless networks: Key techniques and open issues,'' \emph{IEEE
  Commun. Surveys Tuts.}, vol.~21, no.~4, pp. 3072--3108, 2019.

\bibitem{konevcny_basic}
J.~Kone{\v{c}}n{\`y}, H.~B. McMahan, D.~Ramage, and P.~Richt{\'a}rik,
  ``Federated optimization: Distributed machine learning for on-device
  intelligence,'' \emph{arXiv preprint arXiv:1610.02527}, 2016.

\bibitem{McMahan2017}
B.~McMahan, E.~Moore, D.~Ramage, S.~Hampson, and B.~A. y~Arcas,
  ``Communication-efficient learning of deep networks from decentralized
  data,'' in \emph{Artificial intelligence and statistics}.\hskip 1em plus
  0.5em minus 0.4em\relax PMLR, 2017, pp. 1273--1282.

\bibitem{li2020}
T.~Li, A.~K. Sahu, A.~Talwalkar, and V.~Smith, ``Federated learning:
  Challenges, methods, and future directions,'' \emph{IEEE Signal Process.
  Mag.}, vol.~37, no.~3, pp. 50--60, 2020.

\bibitem{bouzinis}
P.~S. Bouzinis, P.~D. Diamantoulakis, and G.~K. Karagiannidis, ``Wireless
  federated learning ({WFL}) for {6G} networks - {P}art {I}: Research
  challenges and future trends,'' \emph{IEEE Commun. Lett.}, pp. 1--1, 2021.

\bibitem{bouzinis2}
------, ``Wireless federated learning ({WFL}) for {6G} networks - {P}art {II}:
  The compute-then-transmit {NOMA} paradigm,'' \emph{IEEE Commun. Lett.},
  vol.~26, no.~1, pp. 8--12, 2022.

\bibitem{lim2020}
W.~Y.~B. Lim, N.~C. Luong, D.~T. Hoang, Y.~Jiao, Y.-C. Liang, Q.~Yang,
  D.~Niyato, and C.~Miao, ``Federated learning in mobile edge networks: A
  comprehensive survey,'' \emph{IEEE Commun. Surveys Tuts.}, vol.~22, no.~3,
  pp. 2031--2063, 2020.

\bibitem{chen2020}
M.~Chen, Z.~Yang, W.~Saad, C.~Yin, H.~V. Poor, and S.~Cui, ``A joint learning
  and communications framework for federated learning over wireless networks,''
  \emph{IEEE Trans. Wireless Commun.}, vol.~20, no.~1, pp. 269--283, 2020.

\bibitem{yang2020}
Z.~Yang, M.~Chen, W.~Saad, C.~S. Hong, and M.~Shikh-Bahaei, ``Energy efficient
  federated learning over wireless communication networks,'' \emph{IEEE Trans.
  Wireless Commun.}, vol.~20, no.~3, pp. 1935--1949, 2020.

\bibitem{chen2020convergence}
M.~Chen, H.~V. Poor, W.~Saad, and S.~Cui, ``Convergence time optimization for
  federated learning over wireless networks,'' \emph{IEEE Trans. Wireless
  Commun.}, vol.~20, no.~4, pp. 2457--2471, 2020.

\bibitem{Wan2021}
S.~Wan, J.~Lu, P.~Fan, Y.~Shao, C.~Peng, and K.~B. Letaief, ``Convergence
  analysis and system design for federated learning over wireless networks,''
  \emph{IEEE J. Sel. Areas Commun.}, vol.~39, no.~12, pp. 3622--3639, 2021.

\bibitem{shi2020}
W.~Shi, S.~Zhou, Z.~Niu, M.~Jiang, and L.~Geng, ``Joint device scheduling and
  resource allocation for latency constrained wireless federated learning,''
  \emph{IEEE Trans. Wireless Commun.}, vol.~20, no.~1, pp. 453--467, 2020.

\bibitem{yang2019}
H.~H. Yang, Z.~Liu, T.~Q. Quek, and H.~V. Poor, ``Scheduling policies for
  federated learning in wireless networks,'' \emph{IEEE Trans. on Commun.},
  vol.~68, no.~1, pp. 317--333, 2019.

\bibitem{konevcny1}
J.~Kone{\v{c}}n{\`y}, H.~B. McMahan, F.~X. Yu, P.~Richt{\'a}rik, A.~T. Suresh,
  and D.~Bacon, ``Federated learning: Strategies for improving communication
  efficiency,'' \emph{arXiv preprint arXiv:1610.05492}, 2016.

\bibitem{fedpaq}
A.~Reisizadeh, A.~Mokhtari, H.~Hassani, A.~Jadbabaie, and R.~Pedarsani,
  ``Fed{PAQ}: A communication-efficient federated learning method with periodic
  averaging and quantization,'' in \emph{International Conference on Artificial
  Intelligence and Statistics}.\hskip 1em plus 0.5em minus 0.4em\relax PMLR,
  2020, pp. 2021--2031.

\bibitem{caldas2018}
S.~Caldas, J.~Kone{\v{c}}ny, H.~B. McMahan, and A.~Talwalkar, ``Expanding the
  reach of federated learning by reducing client resource requirements,''
  \emph{arXiv preprint arXiv:1812.07210}, 2018.

\bibitem{zheng2020}
S.~Zheng, C.~Shen, and X.~Chen, ``Design and analysis of uplink and downlink
  communications for federated learning,'' \emph{IEEE J. Sel. Areas Commun.},
  2020.

\bibitem{amiri2020q}
M.~M. Amiri, D.~Gunduz, S.~R. Kulkarni, and H.~V. Poor, ``Federated learning
  with quantized global model updates,'' \emph{arXiv preprint
  arXiv:2006.10672}, 2020.

\bibitem{chang2020}
W.-T. Chang and R.~Tandon, ``Communication efficient federated learning over
  multiple access channels,'' \emph{arXiv preprint arXiv:2001.08737}, 2020.

\bibitem{wang2021}
Y.~Wang, Y.~Xu, Q.~Shi, and T.-H. Chang, ``Quantized federated learning under
  transmission delay and outage constraints,'' \emph{IEEE J. on Sel. Areas in
  Commun.}, vol.~40, no.~1, pp. 323--341, 2022.

\bibitem{zhu2020}
G.~Zhu, Y.~Du, D.~G{\"u}nd{\"u}z, and K.~Huang, ``One-bit over-the-air
  aggregation for communication-efficient federated edge learning: Design and
  convergence analysis,'' \emph{IEEE Trans. Wireless Commun.}, vol.~20, no.~3,
  pp. 2120--2135, 2020.

\bibitem{shen2021}
C.~Shen, J.~Xu, S.~Zheng, and X.~Chen, ``Resource rationing for wireless
  federated learning: Concept, benefits, and challenges,'' \emph{IEEE Commun.
  Mag.}, vol.~59, no.~5, pp. 82--87, 2021.

\bibitem{tran2019}
N.~H. Tran, W.~Bao, A.~Zomaya, M.~N. Nguyen, and C.~S. Hong, ``Federated
  learning over wireless networks: Optimization model design and analysis,'' in
  \emph{IEEE INFOCOM 2019-IEEE Conference on Computer Communications}.\hskip
  1em plus 0.5em minus 0.4em\relax IEEE, 2019, pp. 1387--1395.

\bibitem{li2019}
X.~Li, K.~Huang, W.~Yang, S.~Wang, and Z.~Zhang, ``On the convergence of fedavg
  on non-iid data,'' \emph{arXiv preprint arXiv:1907.02189}, 2019.

\bibitem{boyd}
S.~Boyd, S.~P. Boyd, and L.~Vandenberghe, \emph{Convex optimization}.\hskip 1em
  plus 0.5em minus 0.4em\relax Cambridge university press, 2004.

\bibitem{lambert}
R.~M. Corless, G.~H. Gonnet, D.~E. Hare, D.~J. Jeffrey, and D.~E. Knuth, ``On
  the {L}ambert {W} function,'' \emph{Adv. Comput. Math.}, vol.~5, no.~1, pp.
  329--359, 1996.

\bibitem{mnist}
Y.~Lecun, C.~Cortes, and C.~Burges, ``Mnist,'' 1998.

\bibitem{adam}
D.~P. Kingma and J.~Ba, ``Adam: A method for stochastic optimization,''
  \emph{arXiv preprint arXiv:1412.6980}, 2014.

\bibitem{amiri2021}
M.~M. Amiri, D.~G{\"u}nd{\"u}z, S.~R. Kulkarni, and H.~V. Poor, ``Convergence
  of update aware device scheduling for federated learning at the wireless
  edge,'' \emph{IEEE Trans. Wireless Commun.}, vol.~20, no.~6, pp. 3643--3658,
  2021.

\end{thebibliography}

\end{document}